\documentclass[a4paper,fleqn,usenatbib]{mnras}

\usepackage{newtxtext,newtxmath}

\usepackage[T1]{fontenc}
\usepackage{ae,aecompl}

\usepackage{graphicx}
\usepackage{amsmath}	
\usepackage{amssymb}	

\usepackage[version=3]{mhchem}

\def\aj{AJ}%
\def\apj{ApJ}%
\def\apjl{ApJ}%
\def\apjs{ApJS}%
\def\ao{Appl.~Opt.}%
\def\aap{A\&A}%
\def\aaps{A\&AS}%
\def\icarus{Icarus}%
\def\mnras{MNRAS}%
\def\prb{Phys.~Rev.~B}%
\def\nat{Nature}%
\title[Optical properties of potential condensates in exoplanetary atmospheres]{Optical properties of potential condensates in exoplanetary atmospheres}

\author[D. Kitzmann et al.]{Daniel Kitzmann,$^1$\thanks{daniel.kitzmann@csh.unibe.ch}
  Kevin Heng$^1$\thanks{kevin.heng@csh.unibe.ch}\\
  $^1$ Universit\"at Bern, Center for Space and Habitability, Gesellschaftsstr. 6, 3012 Bern, Switzerland\\           
}

\date{Accepted 2017 December 1. Received 2017 November 30; in original form 2017 October 13}

\pubyear{2017}

\begin{document}
  \label{firstpage}
  \pagerange{\pageref{firstpage}--\pageref{lastpage}}
  \maketitle

\begin{abstract}
The prevalence of clouds in currently observable exoplanetary atmospheres motivates the compilation and calculation of their optical properties.  First, we present a new open-source Mie scattering code known as \texttt{LX-MIE}, which is able to consider large size parameters ($\sim 10^7$) using a single computational treatment.  We validate \texttt{LX-MIE} against the classical \texttt{MIEV0} code as well as previous studies.  Second, we embark on an expanded survey of the published literature for both the real and imaginary components of the refractive indices of 32 condensate species.  As much as possible, we rely on experimental measurements of the refractive indices and resort to obtaining the real from the imaginary component (or vice versa), via the Kramers-Kronig relation, only in the absence of data.  We use these refractive indices as input for \texttt{LX-MIE} to compute the absorption, scattering and extinction efficiencies of all 32 condensate species.  Finally, we use a three-parameter function to provide convenient fits to the shape of the extinction efficiency curve.  We show that the errors associated with these simple fits in the Wide Field Camera 3 (WFC3), J, H and K wavebands are $\sim 10\%$.  These fits allow for the extinction cross section or opacity of the condensate species to be easily included in retrieval analyses of transmission spectra. We discuss prospects for future experimental work. The compilation of the optical constants and \texttt{LX-MIE} are publicly available as part of the open-source Exoclime Simulation Platform (\url{http://www.exoclime.org}).
\end{abstract}

\nokeywords

\section{Introduction}

Clouds and hazes are expected to be commonplace \citep{Marley2013cctp.book..367M} and have been observationally established to be prevalent in currently observable exoplanetary atmospheres \citep{Pont2008MNRAS.385..109P,Sing2016Natur.529...59S}, with the degree of cloudiness being tentatively correlated with gravity and temperature \citep{Heng2016ApJ...826L..16H,Stevenson2016ApJ...817..141S}.  These findings motivate the inclusion of the effects of these clouds and hazes into retrieval analyses of the spectra of exoplanetary atmospheres.  Cloud and haze formation from first principles is a challenging topic (e.g. \citealt{Helling2008A&A...485..547H}), but once they are formed their effects on the spectrum is a well-understood phenomenon as long as their optical properties are known, because the absorption, scattering and extinction efficiencies may be computed using Mie theory (e.g. \citealt{Hulst1957lssp,Pierrehumbert2010ppc..book.....P}), as we aim to do in the current study.  Effectively, a problem involving particle kinetics embedded in a magnetised fluid reduces to an optics problem.  This reasoning motivates the compilation of a library of refractive indices and the calculation of the corresponding absorption, scattering and extinction efficiencies.

In the current study, we survey the published literature to obtain the measured real and imaginary components of the refractive indices of 32 condensate species. This expands upon the 20 species considered by \citet{Wakeford2015A&A...573A.122W}. A subset of 12 species from the compilation of \citet{Wakeford2015A&A...573A.122W} has recently been used by \citet{Pinhas2017MNRAS.471.4355P} to obtain extinction cross sections of potential condensates and their spectral signatures.  In this work, we also revise some of their used optical constants by using alternative data sources (e.g. for \ce{TiO2}). We especially also include condensates that are expected to form in carbon-rich environments, such as for example \ce{TiC}. 

Additionally we try to cover the entire wavelength range from the ultraviolet to the far infrared for each species, which makes the compilation suitable for atmospheric modelling and simulating spectra.  Heating from the star extends into the ultraviolet (e.g., the Ly$\alpha$ line at 0.12 $\mu$m), and for the purpose of radiative transfer it is essential to compute the extinction of starlight across depth or pressure.  The presence of condensates could alter this extinction profile, which is why it is necessary to have extinction efficiencies to $\sim 0.1$ $\mu$m.  On the other hand, exoplanetary atmospheres emit thermally at peak wavelengths $\sim 1$ $\mu$m $(T/1000 \mbox{ K})$ and also longward of these wavelengths.  To enforce radiative equilibrium requires that one is able to integrate the flux over all wavelengths, down into the long-wavelength tail of the thermal distribution.  The presence of condensates could alter the profile of the flux across wavelength, and it is necessary to be able to compute extinction efficiences to wavelengths as red as $\sim 10$--100 $\mu$m.

We use these refractive indices as input for a newly constructed Mie scattering code, \texttt{LX-MIE}, which we use to compute the absorption, scattering and extinction efficiencies of these 32 condensate species. Based on the reasoning that the extinction efficiency curve has a somewhat general shape (e.g., \citealt{Pierrehumbert2010ppc..book.....P}), we fit a simple three-parameter function to it and quantify the errors associated with the fit in the WFC3, J, H and K wavebands for all 32 species.  These empirical fits allow for the cross section or opacity of these 32 condensate species to be easily included in retrieval analysis of transmission spectra, as was done by \citet{Lee2013ApJ...778...97L}, \citet{Lavie2017AJ....154...91L}, \citet{Oreshenko2017ApJ...847L...3O} and \citet{Tsiaras2017arXiv170405413T}.

In Sect. \ref{sec:mie_theory}, we first briefly review Mie theory as a prelude to describing the improved computational method for our \texttt{LX-MIE} code in Sect. \ref{sec:mie_computations}.  Our compilation and treatment of optical constants are described in Sect. \ref{sec:optical_constants}.  We present the three-parameter fits of the extinction efficiencies in Sect. \ref{sec:analytical_fits}, and finally discuss prospects for future work in Sect. \ref{sec:summary}.

\section{Review of Mie theory}
\label{sec:mie_theory}

To calculate the optical properties of potential condensates we use Mie theory \citep{Mie1908AnP}, assuming a spherical particle shape. For these calculations we constructed a new Mie scattering code that is especially designed to handle situations with large size parameters. Publicly available Mie codes, such as the well-known \texttt{MIEV0} code by \citet{Wiscombe1980ApOpt,Wiscombe1979msca}, are only validated for size parameters of the order of $10^4$. This, for example, is insufficient to treat particles with radii of $1000 \ \mathrm{\mu m}$ at ultraviolet wavelengths. Such large particles are, for example, found in the atmosphere of Earth in the form of ice crystals \citep{SchmittGRL:GRL55211}. 
Large particle sizes can usually be expected if clouds form from the major atmospheric constituent. If such an atmosphere becomes supersaturated, cloud particles can grow quickly due to the huge amount of available, condensible material. For example, \citet{Colaprete2003} estimated that the mean size of \ce{CO2} ice particles in a carbon dioxide-dominated atmosphere of early Mars would be $1000 \ \mathrm{\mu m}$. Other astrophysical environments where large dust particles are known to exist are protoplanetary disks. The dust grains in these disks can easily grow to sizes in the centimetre range \citep{Birnstiel2012A&A...539A.148B, Mordasini2014A&A...572A.118M}.

While geometric optics approximations can be used to determine the optical properties of large particles, it would be preferable to have a Mie code, able to describe the entire wavelength range of interest without needing to switch to different approximations at certain wavelengths and particle sizes. The Mie scattering code described in the following can easily handle size parameters of the order $10^7$ and higher, including the calculation of the scattering and absorption efficiencies, asymmetry parameter, as well as the scattering phase function and its Legendre moments.

Condensates in astrophysical environments are probably not spherical but rather fractal-like structures. However, in-situ measurements of the actual particle shapes in these environments are unavailable. Furthermore, the scattering properties of such non-spherical particles are extremely computationally demanding \citep{Rother2009ewsn.book.....R}. Thus, for simplicity we employ the approximation of spherical particles in the following. Mie theory can also be used to approximate the optical properties of non-spherical or fractal particles. This was done, for example, for fractal haze particles in the atmosphere of the early Earth by \citet{Arney2016AsBio..16..873A} or to describe non-spherical water ice cloud crystals in Earth-like atmospheres by \citet{Kitzmann2010A&A...511A..66K}.

With the assumption of spherical particles the cross-sections for a distinct particle radius $a$ and wavelength $\lambda$ are then given by \citep{Hulst1957lssp,BohrenHuffman1998asls}:
\begin{eqnarray}
  C_\mathrm{ext}(a,\lambda) & = & \frac{2\pi a^2}{x^2} \sum_{n=1}^{\infty} (2n+1) 
  \mathrm{Re} \left( a_{n}(m(\lambda),x) + b_{n}(m(\lambda),x) \right) \ , \nonumber\\ 
  C_\mathrm{sca}(a,\lambda) & = & \frac{2\pi a^2}{x^2} \sum_{n=1}^{\infty} (2n+1) 
  \left( \left| a_{n}(m(\lambda),x) \right|^2 + \left| b_{n}(m(\lambda),x) \right|^2 \right) \ , \nonumber\\
  C_\mathrm{abs}(a,\lambda) & = & C_\mathrm{ext}(a,\lambda) - C_\mathrm{sca}(a,\lambda)
  \label{eq:mie_series}
\end{eqnarray}
with the dimensionless size parameter
\begin{equation}
  x = \frac{2 \pi a}{\lambda} \ .
\end{equation}
The Mie coefficients $a_{n}(m(\lambda),x)$ and $b_{n}(m(\lambda),x)$ are defined with respect to the wavelength-dependent complex refractive index
\begin{equation}
  m(\lambda) = n(\lambda) - \mathrm{i} k(\lambda)
\end{equation}
of the considered material\footnote{In the following we are dropping the $\lambda$ for the simplification of notation.}. For a monodisperse population of condensates, it is convenient to specify the dimensionless efficiencies,
\begin{equation}
  Q = \frac{C}{\pi a^2},
\end{equation}
instead of the corresponding cross-sections.

\subsection{Scattering phase function}

To quantify the role of scattered radiation for e.g. the scattering greenhouse effect of cloud particles in extrasolar planetary atmospheres the corresponding angular distribution of the scattered radiation has to be specified by a scattering phase function $p(a,\alpha)$ which describes the probability of a photon to be scattered with a scattering angle $\alpha$. The Mie intensity functions $i_1(a,\alpha)$ and $i_2(a,\alpha)$ 
\begin{eqnarray}
  i_1(a,\alpha) & = & \left| \sum_{n=1}^\infty \frac{2n+1}{n(n+1)} \left[ a_n(m,x) \pi_n(\alpha) + b_n(m,x) \tau_n(\alpha) \right] \right|^2 \nonumber \ ,\\
  i_2(a,\alpha) & = & \left| \sum_{n=1}^\infty \frac{2n+1}{n(n+1)} \left[ a_n(m,x) \tau_n(\alpha) + b_n(m,x) \pi_n(\alpha) \right] \right|^2
  \label{eq:mie_intensities}
\end{eqnarray}
are used for the description of $p(a,\alpha)$. The angular eigenfunctions $\pi_n(a,\alpha)$ and $\tau_n(a,\alpha)$ are defined as
\begin{equation}
  \pi_n(a,\alpha) = \frac{P_n^1(\cos \alpha)}{\sin \alpha} \quad \mathrm{and} \quad \tau_n(a,\alpha) = \frac{\mathrm{d} P_n^1(\cos \alpha)}{\mathrm{d} \alpha} \ ,
  \label{eq:angular_eigenfunctions}
\end{equation}
where $P_n^j$ are the associated Legendre polynomials. The scattering phase function $p(a,\alpha)$ is given in terms of these intensities by\footnote{We note that the phase function defined here corresponds to \citet{Wiscombe1980ApOpt} but it differs from that given in \citet{Hulst1957lssp} or \citet{BohrenHuffman1998asls} by a factor of $4 \pi$.}:

\begin{equation}
  p(a,\alpha) = \frac{\lambda^2}{2 \pi} \frac{i_1(a,\alpha) + i_2(a,\alpha)}{C_{\mathrm{sca}}(a)} \ .
\end{equation}
It can also be represented as a series of Legendre polynomials \citep{Chandrasekhar1960ratr}
\begin{equation}
  p(a,\alpha) = \sum_{n=0}^{\infty} (2 n + 1) P^0_n(\cos \alpha) \chi_n(a)
\end{equation}
with
\begin{equation}
  \chi_n(a) = \frac{1}{2} \int_{-1}^{1} p(a,\alpha) P^0_n(\cos \alpha) \mathrm d\cos\alpha \ .
  \label{eq:leg_series_moments}
\end{equation}
In practise the phase function expansion series is truncated at $n=N_\mathrm{max}$ where the $\chi_{N_\mathrm{max}}(a)$ becomes numerically zero.

The asymmetry parameter $g(a)$ is another important parameter to characterise the scattering properties of cloud particles. It is defined as the mean cosine of the scattering angle
\begin{equation}
  g(a) = \langle\cos \alpha\rangle = \frac{1}{2} \int_0^\pi p(\alpha) \cos(\alpha) \sin(\alpha) \mathrm d\alpha \ .
\end{equation}
If more light is scattered in forward direction ($\alpha > 90^\circ$), g is positive, while $g<0$ is obtained for dominant scattering in the backward direction ($\alpha > 90^\circ$). For $g=0$ the phase function is symmetric around $\alpha=90^\circ$ (but not necessarily isotropic). Note, that $g$ exactly corresponds to the $\chi_1$ moment of the Legendre series expansion (cf. Eq. (\ref{eq:leg_series_moments})).

\section{An improved Mie code: \texttt{LX-MIE}}
\label{sec:mie_computations}

The optical properties of spherical particles can be determined from Eq. (\ref{eq:mie_series}). The most difficult task is thereby the calculation of the Mie coefficients $a_n$ and $b_n$, given by
\begin{eqnarray}
  a_n(m,x) &=& \frac{ \psi_n(x)\psi'_n(mx) - m\psi'_n(x)\psi_n(mx)}{\zeta_n(x) \psi'_n(mx) - m\zeta'_n(x)\psi_n(mx)} \ , \nonumber\\
  b_n(m,x) &=& \frac{ m\psi_n(x)\psi'_n(mx) - \psi'_n(x)\psi_n(mx)}{m\zeta_n(x) \psi'_n(mx) - \zeta'_n(x)\psi_n(mx)}
  \label{eq:mie_coefficients}
\end{eqnarray}
with the Riccati-Bessel functions $\zeta_n$, $\psi_n$ and their corresponding derivatives \citep{BohrenHuffman1998asls}. In the following we first briefly summarise the standard approach for the calculation of $a_n$ and $b_n$. However, under certain conditions particles might reach very large size parameters of the order of $10^5$ which cannot be handled by the standard Mie calculation procedure. Consequently, we developed a new modified Mie-program code \texttt{LX-MIE}, especially suited for large size parameters $x$. The improvements we introduced in these Mie calculations are discussed in Sect. \ref{sec:improved_mie}.

\subsection{Standard procedure for Mie calculations}

The calculation of the Mie coefficients present several computational challenges. The most important one is the calculation of $\psi'_n(mx)$ and $\psi_n(mx)$ which can be solved by introducing the logarithmic derivative $A_n(x)$ \citep{Infeld1947QAM}
\begin{equation}
  A_n(mx) = \frac{\mathrm{d}(\log \psi_n(mx))}{\mathrm{d}(mx)} = \frac{\psi'_n(mx)}{\psi_n(mx)} = -\frac{n}{mx} + \frac{J_{n-1/2}(mx)}{J_{n+1/2}(mx)}
\end{equation}
where $J_{n\pm1/2}$ are the Bessel functions of the first kind.
Using the logarithmic derivative, the Mie coefficients can be re-written in the following form
\begin{eqnarray}
  a_n(m,x) &=& \frac{(A_n(mx)/m + n/x)\psi_n(x) - \psi_{n-1}(x)}{(A_n(mx)/m + n/x)\zeta_n(x) - \zeta_{n-1}(x)} \ , \nonumber\\
  b_n(m,x) &=& \frac{(mA_n(mx) + n/x)\psi_n(x) - \psi_{n-1}(x)}{(mA_n(mx) + n/x)\zeta_n(x) - \zeta_{n-1}(x)} \ .
\end{eqnarray}
The logarithmic derivative $A_n(mx)$ can then be obtained by an upward or a downward recursion. The upward recursion is only stable for special cases (e.g. if $\mathrm{Im}(m)$ is small), while the downward recursion
\begin{equation}
  A_{n-1}(mx) = \frac{n}{mx} - \frac{1}{\frac{n}{mx} + A_n(mx)}
  \label{eq:log_deriv_rec}
\end{equation}
is always stable \citep{Wiscombe1979msca}. For a downward recursion, however, the initial value $A_{N}(mx)$ must be known a priori, where $n=N$ is the last term to be taken into account in the (in principle infinite) Mie series. Based on numerical studies with size parameters of up to 20\,000, \citet{Wiscombe1980ApOpt} suggested the following values:

\begin{equation}
  N = \tilde{N}(x) = \left\{ 
  \begin{array}{ll} 
    x + 4 x^{1/3} + 1 & \quad 0.02 \leq x \leq 8\\ 
    x + 4.05 x^{1/3} + 2 & \quad 8 < x < 4200\\ 
    x + 4 x^{1/3} + 2 & \quad 4200 \leq x \leq 20\,000
  \end{array} 
  \right.
  \label{eq:Wiscombe_N}
\end{equation}
This commonly used criterion for the value of $N=\tilde{N}(x)$ from \citet{Wiscombe1980ApOpt} was derived empirically for $x\leq 20\,000$, but lacks a clear mathematical justification.
The corresponding initial value $A_N(mx)$ for the recurrence relation is usually obtained by the continuous fraction method introduced by \citet{Lentz1976ApOpt}\footnote{We note that in principle $A_n(mx)$ can be calculated with the continuous fractions from \citet{Lentz1976ApOpt} for any value of $n$, but the recurrence relation Eq. (\ref{eq:log_deriv_rec}) is computationally much faster.}. The Riccati-Bessel functions $\zeta_n(x)$ and $\psi_n(x)=\mathrm{Re} \, \zeta_n(x)$ are calculated by an upward recursion \citep{AbramowitzStegun1972}, i.e.
\begin{equation}
  \zeta_{n+1}(x) = \frac{2n + 1}{x} \zeta_n(x) - \zeta_{n-1}(x) \ .
\end{equation}
This recursion, however, is not always stable, especially if $n \gg x$ numerical errors start to dominate the calculations making the upward recursion unstable. 

\subsection{Improved Mie calculations for large size parameters}
\label{sec:improved_mie}

To perform Mie calculations for large size parameters the method to calculate the Riccati-Bessel functions in the Mie coefficients has to be improved. Therefore, following \citet{Shen2005PIERS} we write the Mie coefficients $a_n$ and $b_n$ in the form
\begin{eqnarray}
  a_n(m,x) &=& B_n(x) \frac{A_n(mx)/m - A_n(x)}{A_n(mx)/m - C_n(x)} \ , \nonumber \\
  b_n(m,x) &=& B_n(x) \frac{A_n(mx) m - A_n(x)}{A_n(mx) m - C_n(x)}
  \label{eq:mie_coeff_ratios}
\end{eqnarray}
with
\begin{eqnarray}
  B_n(x) &=& \frac{\psi_n(x)}{\zeta_n(x)} \nonumber \ ,\\
  C_n(x) &=& \frac{\zeta'_n(x)}{\zeta_n(x)} \ .
\end{eqnarray}
In this formulation, the Mie coefficients are only functions of ratios of Riccati-Bessel functions and their derivatives. Thus, the direct calculation of single Riccati-Bessel functions is completely avoided. The computation of these ratios is numerically much more stable than that of single Riccati-Bessel functions which results in higher accuracy and computational stability for large $x$ and $n$. Therefore, large size parameters of e.g. $x=10^6$ can be used without numerical problems in our Mie calculations.

The ratios $B_n(x)$ and $C_n(x)$ also obey recurrence relations, namely:
\begin{eqnarray}
  B_n(x) &=& B_{n-1}(x) \frac{C_n(x)+n/x}{A_n(x) + n/x} \nonumber \ , \\
  C_n(x) &=& -\frac{n}{x} + \frac{1}{\frac{n}{x} - C_{n-1}(x)} \ .
\end{eqnarray}
Both ratios are calculated here via upward recursions, starting with:
\begin{eqnarray}
  B_1(x) &=& \left( 1 + \mathrm{i} \frac{\cos x + x \sin x}{\sin x - x \cos x} \right)^{-1} \nonumber \quad \mathrm{and} \\
  C_0(x) &=& -\mathrm{i} \ .
\end{eqnarray}
The coefficients $A_n(mx)$ and $A_n(x)$ are again calculated via the downward recurrence Eq. (\ref{eq:log_deriv_rec}) and its initial values $A_N(mx)$ and $A_N(x)$ are obtained from the continuous fraction method after \citet{Lentz1976ApOpt}.

To obtain a non-empirical criterion for the value of $N=N(x)$ in the case of large $x$, we adopt the method published by \citet{Cachorro1991}. They proved mathematically that the convergence properties of the Mie series depend almost completely on the ratios $B_n(x)$ in Eq. (\ref{eq:mie_coeff_ratios}). Based on their findings they concluded that the truncation criterion for the series (i.e. the value of $N$) as a function of the desired accuracy $\epsilon$ is given by
\begin{equation}
  |\mathrm{Im} \, \zeta_N(x) | \geq \sqrt{\epsilon^{-1}} \ ,
\end{equation}
considering the fact $\mathrm{Re}\,\zeta_n(x) = \psi_n(x)$ is negligible compared to $\mathrm{Im} \, \zeta_n(x)$ for large values of $n$. For their general relation of the truncation criterion
\begin{equation}
  N=\tilde{N}(x) = x + c \cdot x^{1/3}
\end{equation}
they derived the constants $c$ for several values of $\epsilon$. For our Mie calculations, we adopt their value of $c=4.3$ which corresponds to $\epsilon = 10^{-8}$.

However, both the criteria of \citet{Wiscombe1980ApOpt} and of \citet{Cachorro1991} are only functions of the size parameter $x$ and do not depend on the refractive index $m$. As discussed by \citet{Cachorro1991} the conclusion of $\tilde{N}$ being only a function of $x$ is not universally true. In extremely rare cases resonant terms can exist in the Mie series (depending on $m$) which invalidates the assumption that $\tilde{N}$ is a function of $x$ only. In order for these resonance effects to appear in the Mie series, $x$ and $m$ would have to be fine tuned with very high precision \citep{Cachorro1991}. Therefore, it is unlikely to encounter such resonances in practical Mie calculations by coincidence.

With given Mie coefficients the remaining calculations are straightforward. For the Mie intensities (Eq. [\ref{eq:mie_intensities}]) the angular eigenfunctions $\pi_n(a,\alpha)$ and $\tau_n(a,\alpha)$ can also be written as recurrence relations which are calculated via upward recursion (see \citet{Wiscombe1979msca} for details). To calculate the moments of the Legendre series expansion of the scattering phase function (Eq. (\ref{eq:leg_series_moments})) the procedure from \citet{Dave1970ApOpt} can be used. The newly developed Mie scattering code \texttt{LX-MIE} has been applied to various test examples for validation (see Appendix \ref{sec:appendix_validation}).

\section{Optical constants}
\label{sec:optical_constants}

In this section, we describe a new collection of optical constants, the real part $n$ and the imaginary part $k$ of the refractive index, for a broad set of potential condensates expected in extrasolar gas planets. 

Previously, a set of optical constants for 20 species was published by \citet{Wakeford2015A&A...573A.122W}. We here extend and revise their published optical constants by including more dust species and exploring other data sources to obtain refractive indices over a broad wavelength range from the ultraviolet to the far infrared. The only two species that have the same optical constants in \citet{Wakeford2015A&A...573A.122W} and this publication are \ce{KCl} and \ce{NaCl}. 
We note that the data set for the refractive index of \ce{TiO2} \citep{kangarloo2010a,kangarloo2010b} used by \citet{Wakeford2015A&A...573A.122W} has been retracted by the corresponding journal and, thus, requires revision.
We especially focus on amorphous condensates which are probably the more common form of dust particles in astrophysical environments. 
In addition to  \citet{Wakeford2015A&A...573A.122W}, we also take anisotropic crystals and temperature-dependent refractive indices into account, if the corresponding data is available.

\begin{figure*}
	\includegraphics[width=155mm]{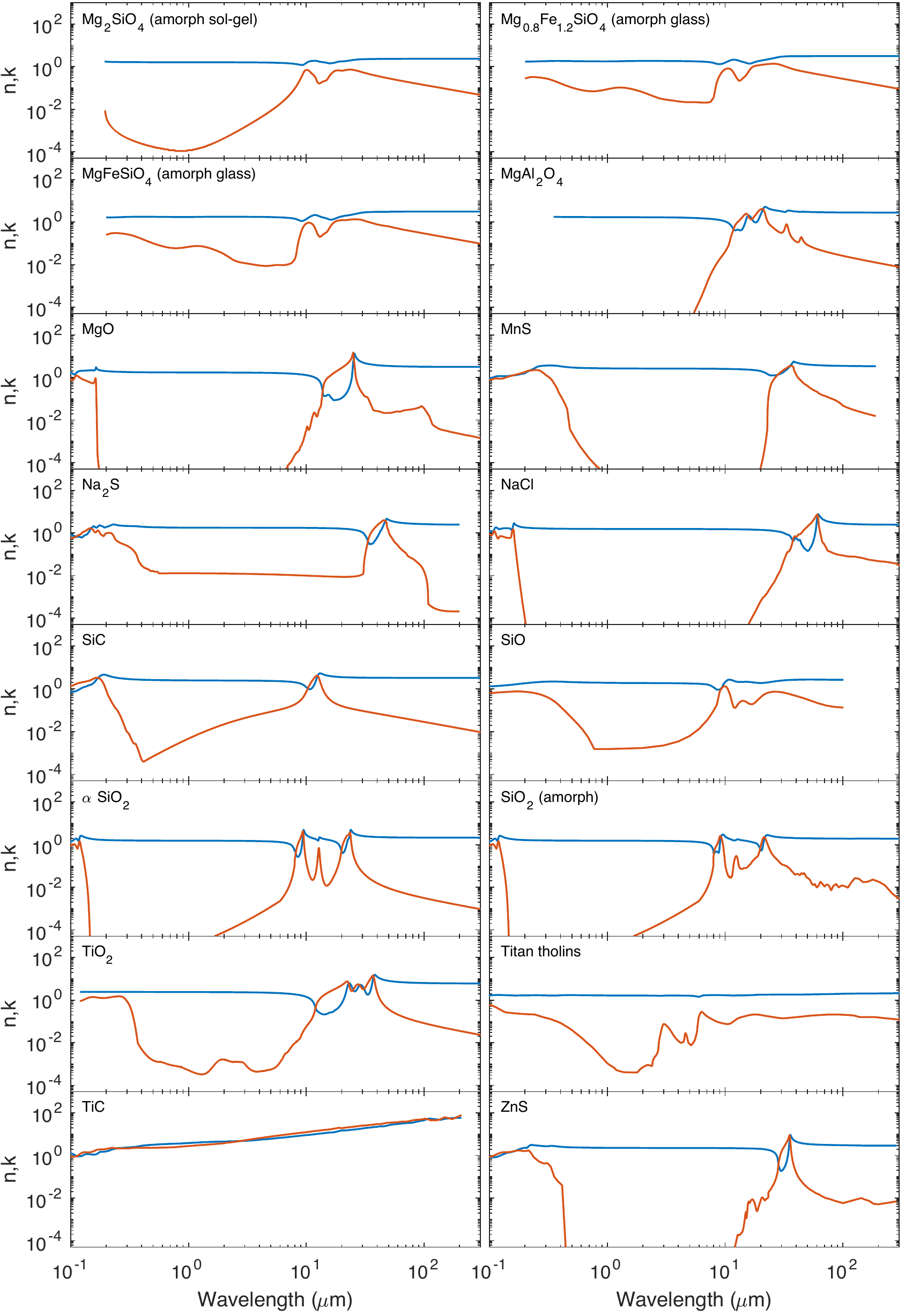}
	\caption{Optical constants of a set of potential condensates. The plots show the real ($n$, blue) and imaginary ($k$, red) parts of the refractive index. The optical constants have been compiled from several different sources. Details on the compilation are given in Sect. \ref{sec:optical_constants} and Table \ref{tab:optical_constants}.}
	\label{fig:optical_constants}
\end{figure*}

Our list includes several common silicates, oxides, sulfides, chlorides, but also carbon bearing condensates. We also include some solid solutions of the olivine and pyroxene groups. Most of the optical constants are given from the ultraviolet to the far-infrared range of wavelengths, which makes them suitable to be included in atmospheric models or to study their impact on spectra. The optical constants have been obtained from several sources and are listed in Table \ref{tab:optical_constants}. The plots of the $n$ and $k$ values as a function of wavelength for all considered condensates are shown in Fig. \ref{fig:optical_constants}. For enstatite and quartz we respectively include two different data sets. In case of \ce{MgSiO3} these are two different amorphous states and for \ce{SiO2} we include its $\alpha$-crystal as well as amorphous form.

In theory, $n$ and $k$ are not independent but as the real and imaginary part of a complex number must follow the Kramers-Kronig relations \citep{Kronig1926JOSA, Kramers1927}
\begin{equation}
n(\omega) - 1 = \frac{2}{\pi} P \int_0^\infty \frac{\omega' k(\omega')}{\omega'^2 - \omega^2} \mathrm d \omega' 
\end{equation}
\begin{equation}
k(\omega) = - \frac{2\omega}{\pi} P \int_0^\infty \frac{n(\omega') - 1}{\omega'^2 - \omega^2} \mathrm d \omega' \ ,
\end{equation}
where $\omega = 2 \pi c / \lambda$ is the angular frequency and $P$ denotes the principal value of the singular integral.
Theoretically, all $n$ and $k$ values shown in Fig. \ref{fig:optical_constants} must satisfy these two relations. In practice, however, this is not always the case. Since most of the data is compiled from different sources, with diverse experimental setups, measurement errors, etc., it is usually impossible to satisfy the Kramers-Kronig relations over the entire wavelength range. While it would be possible to iterate $n$ and $k$ via the Kramers-Kronig relations to obtain a consistent set of optical constants, the usefulness of this approach is limited by the fact that the resulting refractive index would not be consistent anymore with the measured experimental data. Since we do not wish to judge each experimentally derived refractive index on the basis of their experimental setup or measurement accuracy, we choose not to perform an iteration here.

Furthermore, the calculation of $k$ in the large transparency windows, where it attains values of smaller than say $10^{-5}$ would mean that $n$ would have to be measured with a comparable accuracy in order to produce reliable $k$-values via the Kramers-Kronig relations. Experimental measurement errors, however, are well above this required accuracy. 

Gaps in the experimental data are closed by extrapolation of either $n$ or $k$ and then using the Kramers-Kronig relations to obtain the other, missing part of the refractive index over the wavelength range of the data gap. For this procedure we employ the singly-subtractive Kramers-Kronig relations that use experimental data at other wavelengths to constrain the missing coefficients in the data gap \citep{Lucarini2005kkro.book.....L}.

In case of \ce{MnS}, no infrared data was available, which includes in particular the sulphur feature. Here, we follow the approach of \citet{Morley2012ApJ...756..172M} and use extrapolations based on the other two sulphur-bearing species \ce{Na2S} and \ce{ZnS} to reconstruct the missing part. For troilite we combine the two data sets of  \citet{Pollack1994ApJ...421..615P} and \citet{Henning1997AA...327..743H}. The compilation of \citet{Pollack1994ApJ...421..615P} used a few experimental data points in combination with extrapolations to account for the missing data. Especially in the infrared region, this set is not consistent with the measured data provided by \citet{Henning1997AA...327..743H}. We therefore replace the IR data with the latter one and use the Kramers-Kronig relations to consistently combine the IR part with the measured shortwave data provided in \citet{Pollack1994ApJ...421..615P}. This leads to a shift of the strong iron feature near 3 $\micron$ to slightly larger wavelengths compared to the original compilation provided by \citet{Pollack1994ApJ...421..615P}.

Some condensates, such as graphite, are anisotropic, i.e. their optical properties depend on the direction of the incident light relative to the crystals' principal axes. In case directional-dependent data was available for these condensates, we calculate a mean refractive index by weighting the dielectric functions in each direction by $1/3$ and converting them to the corresponding $n$ and $k$ values. In Table \ref{tab:optical_constants}, these materials are marked as \textit{anisotropic}. 

In case temperature-dependent data is available, we always choose a temperature close to the ones expected in atmospheres of brown dwarfs or giant exoplanet where the condensates are expected to form. In most cases, though, the optical constants have been measured at room temperature.

\begin{table*}
	\caption{Collection of optical constants for a set of potential condensates in atmospheres of extrasolar planets and brown dwarfs. The Comments column lists additional information on temperature-dependent data (if available), and amorphous or anisotropic materials. If no temperature information is given, measurements where either performed at room temperature or no information is provided in the corresponding publications.}
	\label{tab:optical_constants}
	\begin{tabular}{llllll}
		\hline
		Condensate   & Name & Wavelength range & Comments & \multicolumn{2}{c}{Data Source}\\
		             &      & \multicolumn{1}{c}{\micron} &        & Reference  & Wavelength range\\ 
		             &      &                             &        &            & \multicolumn{1}{c}{\micron}\\
		\hline
		\ce{Al2O3}   & corundum   & 0.2 - 500 & amorphous; 873 K            & \citet{Begemann1997ApJ...476..199B}$^*$$^t$ & 7.8 - 500  \\
	                 &            &           &           & \citet{Koike1995Icar..114..203K}$^t$    & 0.2 - 12.0  \\[2pt]
		\ce{C}       & graphite   & $10^{-4} - 10^{5}$ & anisotropic & \citet{Draine2003ApJ...598.1017D}$^t$   & $10^{-4} - 10^{5}$ \\[2pt]
		\ce{CaTiO3}  & perovskite & 0.035 - 5.8 $\cdot 10^{5}$ & & \citet{Posch2003ApJS..149..437P}$^t$ & $2 - 5.8 \cdot 10^{5}$ \\
		             &            &                     & & \citet{Ueda1998JPCM...10.3669U}$^f$ & 0.02 - 2.0   \\[2pt]
		\ce{Cr}      &            & 0.04 - 500 &  & Lynch\&Hunter in \citet{Palik1991hocs.book}$^t$  &  0.04 - 31   \\
		             &            &                                      &  & \citet{Rakic:98}$^t$  & 31 - 500 \\[2pt]
		\ce{Fe}[s]   &            & $1.2 \cdot 10^{-4}$ - 286 &           & Lynch\&Hunter 
		in \citet{Palik1991hocs.book}$^t$  & $1.2 \cdot 10^{-4}$ - 286 \\[2pt] 
		\ce{Fe2O3}   & hematite   & 0.1 - 1000 & &   A.H.M.J. Triaud$^*$$^t$                & 0.1 - 1000  \\[2pt]
		\ce{Fe2SiO4} & fayalite   & 0.4 - $10^{4}$ & anisotropic & \citet{Fabian2001AA...378..228F}$^*$$^t$  & 0.4 - $10^{4}$ \\[2pt]
		\ce{FeO}     & w\"ustite  & 0.1 - 500 &  & \citet{Henning1995AAS..112..143H}$^*$$^t$ & 0.1 - 50  \\[2pt]
		\ce{FeS}     & troilite   & 0.1 - 487 &  & \citet{Pollack1994ApJ...421..615P}$^t$  & 0.1 - $10^{4}$ \\
		             &            &           &  & \citet{Henning1997AA...327..743H}$^t$   & 20 - 487\\[2pt]
		\ce{H2O}[s]  &            & 0.04 - 2 $\cdot 10^{6}$ & & \citet{Warren1984ApOpt..23.1206W}$^t$  & 0.04 - 2 $\cdot 10^{6}$\\[2pt]
		\ce{KCl}     & sylvite    & 0.03 - $10^{3}$ &  & Palik in \citet{Palik1985hocs.book}$^t$ & 0.03 - $10^{3}$ \\[2pt]
		\ce{Mg2SiO4} & forsterite & 0.2 - 950 & amorphous (sol-gel)  & \citet{Jager2003AA...408..193J}$^*$$^t$ & 0.2 - 950 \\[2pt]
		\ce{Mg_{0.8}Fe_{1.2}SiO4} & olivine    & 0.2 - 500 & amorphous (glassy) & \citet{Dorschner1995AA...300..503D}$^*$$^t$ & 0.2 - 500 \\[2pt]        
		\ce{MgFeSiO4}& olivine    & 0.2 - 500  & amorphous (glassy) & \citet{Dorschner1995AA...300..503D}$^*$$^t$ & 0.2 - 500\\[2pt]
		\ce{MgSiO3}  & enstatite  & 0.2 - 500  & amorphous (glassy) & \citet{Dorschner1995AA...300..503D}$^*$$^t$   & 0.2 - 500\\[2pt]
		\ce{MgSiO3}  & enstatite  & 0.2 - 1000 & amorphous (sol-gel) & \citet{Jager2003AA...408..193J}$^*$$^t$ & 0.2 - 1000\\[2pt]                   
		\ce{Mg_{0.4}Fe_{0.6}SiO3} & pyroxene & 0.2 - 500 & amorphous (glassy) & \citet{Dorschner1995AA...300..503D}$^*$$^t$ & 0.2 - 500\\[2pt] 		\ce{Mg_{0.5}Fe_{0.5}SiO3} & pyroxene & 0.2 - 500 & amorphous (glassy) & \citet{Dorschner1995AA...300..503D}$^t$ & 0.2 - 500\\[2pt] 
		\ce{Mg_{0.8}Fe_{0.2}SiO3} & pyroxene & 0.2 - 500 & amorphous (glassy) & \citet{Dorschner1995AA...300..503D}$^*$$^t$ & 0.2 - 500\\[2pt] 
		\ce{MgAl2O4} & spinel     & 0.35 - $10^{4}$ & 928 K & \citet{Zeidler2011AA...526A..68Z}$^*$$^t$ & 6.6 - $10^{4}$\\
		             &            &  &                    & Tropf \& Thomas in \citet{Palik1991hocs.book}$^t$ & 0.35 - 11.9\\[2pt]  
		\ce{MgO}     & periclase  & 0.16 - 625 & & Roessler \& Huffman in \citet{Palik1991hocs.book}$^t$ & 0.16 - 625\\[2pt]  
		\ce{MnS}     &            & 0.09 - 190 & & \citet{Huffman1967PhRv..156..989H}$^{tf}$ & 0.05 - 13\\
		             &            &            & & \citet{Montaner1979PSSAR..52..597M}$^{f}$ & 2.5 - 200\\[2pt]
		\ce{Na2S}    &            & 0.04 - 200 & & \citet{Khachai2009JPCM...21i5404K}$^{f}$  & 0.04 - 63\\
		             &            &            & & \citet{Montaner1979PSSAR..52..597M}$^{f}$ & 2.5 - 200\\[2pt]           
		\ce{NaCl}    &            & 0.05 - $ 3 \cdot 10^{4}$ & & Eldrige \& Palik in \citet{Palik1985hocs.book}$^t$ & 0.05 - $ 3 \cdot 10^{4}$\\[2pt]
		\ce{SiC}     & moissanite & $10^{-3} - 10^{3}$ &  & \citet{Laor1993ApJ...402..441L}$^t$ & $10^{-3} - 10^{3}$\\[2pt]
		\ce{SiO}     &            & $0.05 - 100$ & & Philipp in \citet{Palik1985hocs.book}$^t$ & $0.05 - 100$\\[2pt]
		\ce{SiO2}    & quartz     & $0.05 - 10^{4}$ & $\alpha$-crystal; 928 K;  & \citet{Zeidler2013AA...553A..81Z}$^*$$^t$ & 6.25 - $10^{4}$\\
		             &            &  & anisotropic                & Philipp in \citet{Palik1985hocs.book}$^t$ & 0.05 - 8.4\\[2pt]
		\ce{SiO2}    & quartz     & 0.05 - 490 & amorphous; 300 K &  \citet{Henning1997AA...327..743H}$^*$$^t$ & 6.67 - 490\\
		             &            &   &              & Philipp in \citet{Palik1985hocs.book}$^t$ & 0.05 - 8.4\\[2pt]
		\ce{TiC}     &            & 0.02 - 207 &     & \citet{Koide1990PhRvB..42.4979K}$^t$  &9 $\cdot 10^{-3}$ - 0.9\\
		             &            & &                & \citet{Henning2001AcSpe..57..815H}$^f$ & $10^{-4}$ - 207\\[2pt] 
		\ce{TiO2}    & anatase    & 0.1 - $ 6 \cdot 10^{3}$ & anisotropic  & \citet{Zeidler2011AA...526A..68Z}$^*$$^t$ & 0.4 - 10\\
		             &            & &                            & \citet{Posch2003ApJS..149..437P}$^*$$^t$ & 10 - $ 6 \cdot 10^{3}$\\
		             &            & &                            & \citet{SiefkeADOM201600250}$^t$ & 0.1 -125\\[2pt] 
		\ce{ZnS}     &            & 0.02 - 286 & & Palik \& Addamiano in \citet{Palik1985hocs.book}$^t$ & 0.02 - 286\\[2pt]
		Titan tholins&            & 0.02 - 920 & & \citet{Khare1984Icar...60..127K}$^t$ & 0.02 - 920\\[2pt]      
		\hline
	\end{tabular}\\
	\vspace{1ex}
	\raggedright \textbf{Notes.} $^*$ Optical constants from the Database of Optical Constants for Cosmic Dust, Laboratory Astrophysics Group of the AIU Jena (\url{http://www.astro.uni-jena.de/Laboratory/OCDB/index.html}).\\ 
	$^t$ Data from a printed table or in digital format.\\ 
	$^f$ Data available as figure.
\end{table*}

\begin{figure*}
	\includegraphics[width=155mm]{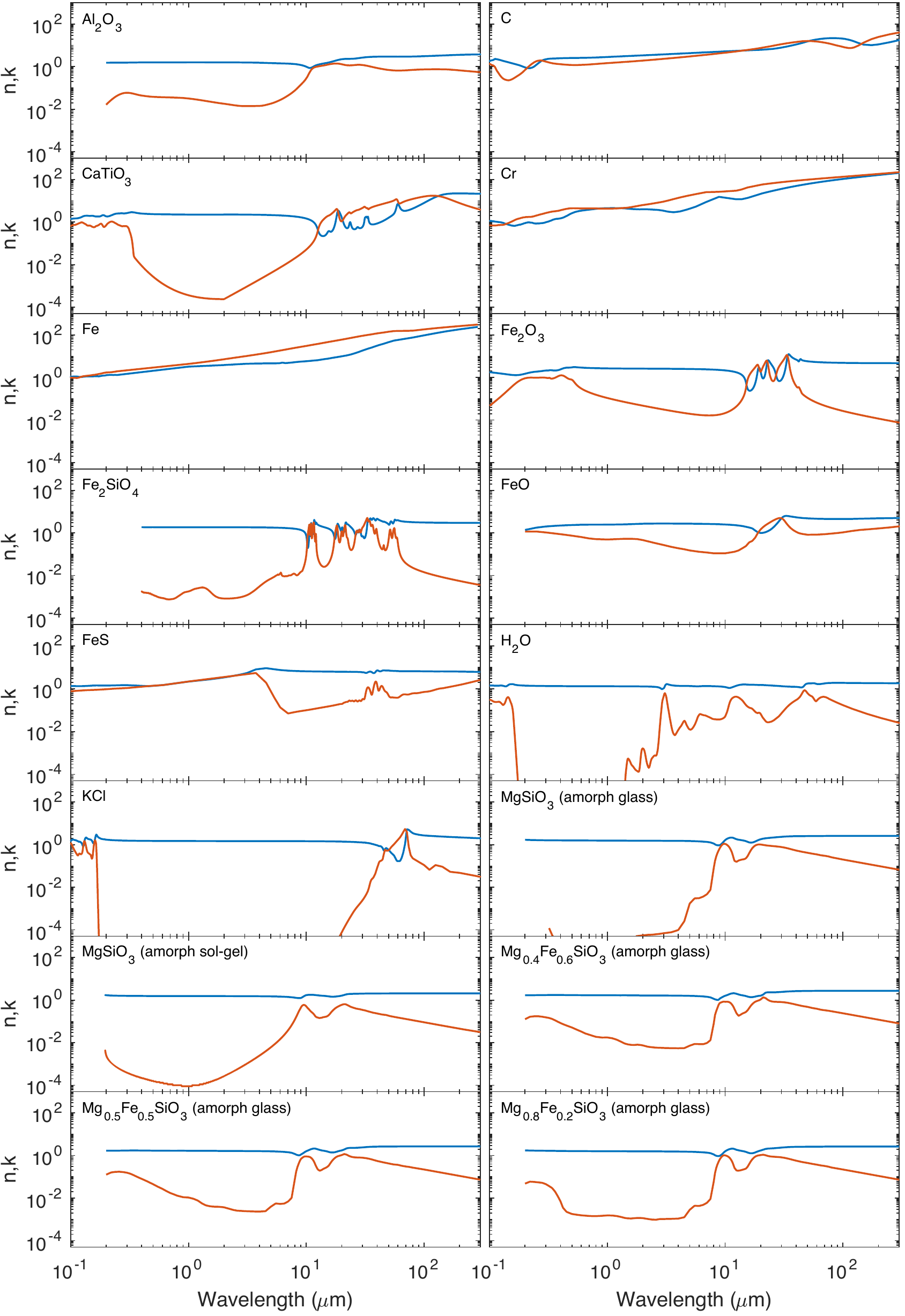}
\end{figure*}

\section{Analytical fits of extinction efficiencies}
\label{sec:analytical_fits}

In retrieval analyses of spectra of exoplanetary atmospheres, there is always a balance between the realism and sophistication of the model, the computational efficiency, and the quality and number of data points.  Including a full-blown, first-principles cloud model in atmospheric retrievals of current spectra is currently unwarranted, given the large number of free parameters involved and the $\sim 10 - 100$ data points typically available for each object.  Instead of including a formation model of clouds in the retrieval, several past efforts have focused on how the clouds would affect the spectrum if assumed to be present \citep{Lee2013ApJ...778...97L, Lavie2017AJ....154...91L, Oreshenko2017ApJ...847L...3O, Tsiaras2017arXiv170405413T}.  Specifically, these studies follow the reasoning laid out in, e.g. \cite{Pierrehumbert2010ppc..book.....P}, that the curves of extinction efficiencies follow a roughly universal shape, somewhat insensitive to the composition of the condensate.  (We will see later that this statement is only partially correct.)  The efficiencies for small and large values of the size parameter ($x$) converge to simpler, analytical expressions.  In other words, our motivation is to provide a library of more accurate fits to augment the approach that is already being pursued in the published literature.

For large values of $x$, i.e. if the particle radius is large compared to the wavelength under consideration, Mie theory yields the large-particle limit, which is given by the simple expressions
\begin{equation}
\lim_{x\rightarrow \infty} Q_\mathrm{abs} = 1 - Q_\mathrm{refl}  \ ,
\end{equation}
\begin{equation}
\lim_{x\rightarrow \infty} Q_\mathrm{sca} = 1 + Q_\mathrm{refl}
\end{equation}
and
\begin{equation}
\lim_{x\rightarrow \infty} Q_\mathrm{ext} = 2 \ .
\end{equation}
The reflection efficiency $Q_\mathrm{refl}$ can be determined by geometric optics calculations (see \citet{BohrenHuffman1998asls} for details).

For particles much smaller than the considered wavelength, i.e. $x = 2 \pi a / \lambda \ll 1 $, Mie theory converges to the limit of Rayleigh scattering. The efficiencies are then given by:
\begin{equation}
  Q_{\mathrm{abs}} = 4x \Im\left\lbrace \frac{m(\lambda)^2 - 1}{m(\lambda)^2 + 2} \right\rbrace
                     \left[ 1 - \frac{4x^3}{3} \Im \left\lbrace \frac{m(\lambda)^2-1}{m(\lambda)^2 + 2} \right\rbrace^2 \right] \ ,
\end{equation}
\begin{equation}
  Q_{\mathrm{sca}} = \frac{8}{3} x^4 \Re \left\lbrace \frac{m(\lambda)^2 - 1}{m(\lambda)^2 + 
  	2} \right\rbrace^2 \ ,
\end{equation}
and
\begin{equation}
Q_{\mathrm{ext}} = Q_{\mathrm{abs}} + Q_{\mathrm{sca}} \ .
\end{equation}

\begin{figure}
	\includegraphics[width=\columnwidth]{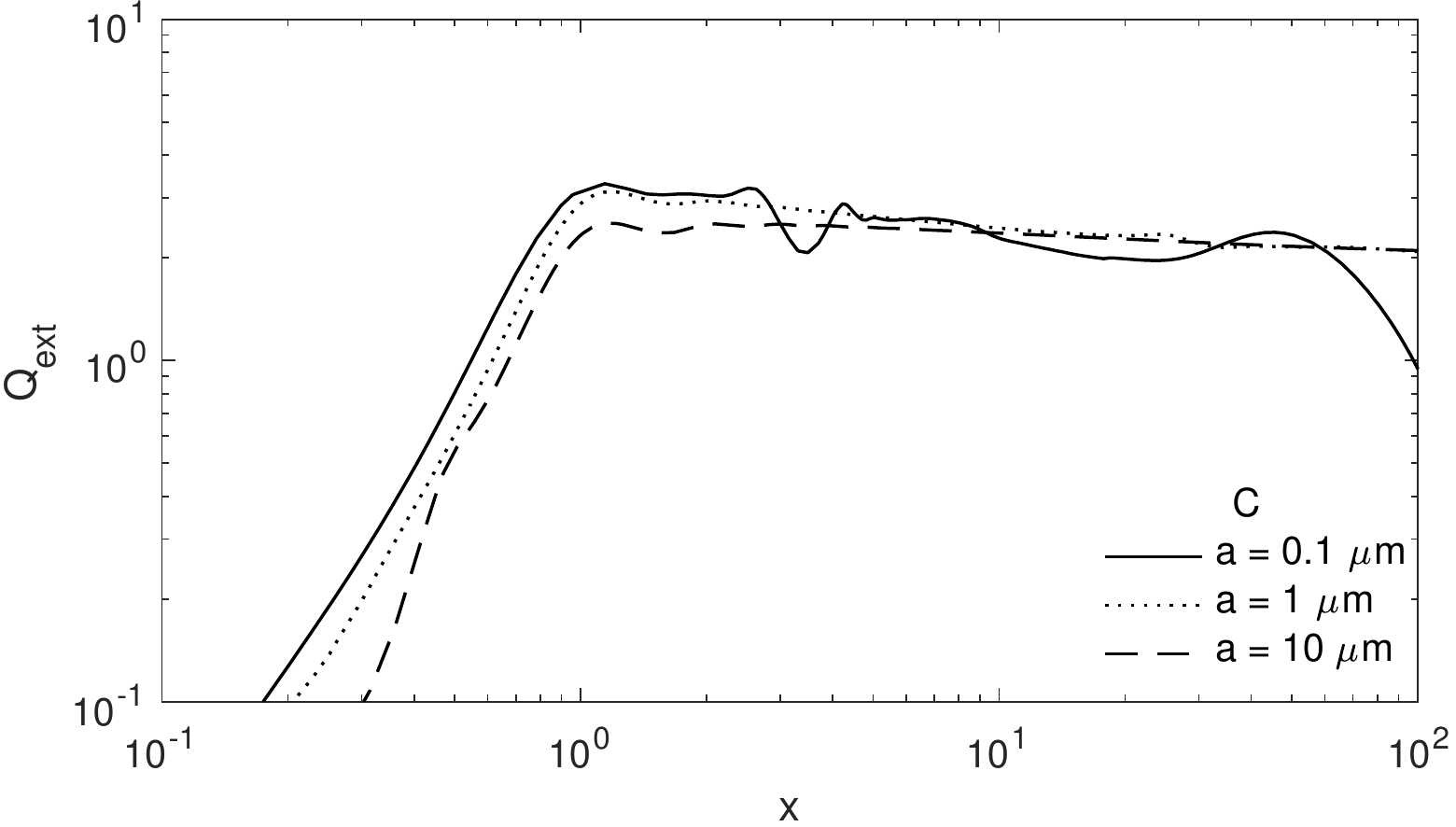}
	\caption{Extinction efficiency of graphite as a function of the size parameter $x$. Shown are the results for three different particle sizes: 0.1 $\micron$ (solid), 1 $\micron$ (dotted), and 10 $\micron$ (dashed).}
	\label{fig:q_ext_graphite}
\end{figure}

An example of the extinction efficiencies of graphite for three different particle sizes is shown in Fig. \ref{fig:q_ext_graphite}. The figure illustrates nicely the two different limits of Mie theory just discussed. Note, however, that the curve for the particles with a radius of 0.1 $\micron$ slightly deviates from the large particle limit due to a drop in the absorptivity of graphite at around a wavelength of $0.1 \micron$ (see Fig. \ref{fig:optical_constants}). The overall shape of the curve can in principle be represented by an almost constant value at large $x$, a slope for small size parameters, and a specific $x$-value where the $Q_{\text{ext}}$ peaks, connecting the two limits.

Given the general shape of the extinction efficiency curve, \citet{Lee2013ApJ...778...97L} proposed a simple analytical fit to describe the overall behaviour of $Q_{\text{ext}}$ as a function of the the size parameter $x$
\begin{equation}
Q_{\text{ext}} = \frac{5}{Q_0 x^{-4} + x^{0.2}} \ .
\end{equation}

This fit is not expected to provide an accurate representation of the extinction efficiencies but can only approximate the overall shape of $Q_{\text{ext}}$. The free parameter $Q_0$ describes the $x$-value where the extinction efficiencies peak. The parameter depends on the condensate optical properties, and thus, on the particle composition. It could therefore be used for a first-order characterisation of condensates in exoplanet transmission spectra, for example. These transmission spectra depend mostly on the extinction properties and not on the detailed contributions of scattering and absorption. The idea of such an analytical fit is, thus, to offer simple retrievable parameters for transmission spectroscopy that would allow for a \textit{basic} characterisation of the condensates present in the planet's atmosphere to a certain degree. It will not be accurate enough to distinguish between different members of e.g. the olivine or pyroxene family, though.

In the following subsection we present our fitting procedure. The corresponding fits and associated errors with respect to the actual $Q_{\text{ext}}$ are discussed in Sect. \ref{sec:fit_results}. 

\subsection{Description of the analytical fit}

We follow the analytical fit of $Q_{\text{ext}}$ as a function of $x$ proposed in \citet{Lee2013ApJ...778...97L} but increase the number of free parameters to three:
\begin{equation}
 Q_{\text{ext}} = \frac{Q_1}{Q_0 x^{-a} + x^{0.2}} \ ,
\end{equation}
where $Q_1$ is a normalisation constant, $Q_0$ determines the $x$-value where $Q_{\text{ext}}$ peaks and $a$ is the slope in the small particle limit.
It is quite obvious that a simple, analytical equation as this cannot describe all the usually complex behaviour of the extinction efficiencies over a large wavelength range, especially the various specific absorption features of the different condensates. We may, however, hope to use the analytic expression in restricted wavelength bands associated with important observational filters.

Here, we focus especially on the WFC3 instrument with the G141 grism and the standard filters $J$, $H$, and $K$ that are used in exoplanet transmission or brown dwarf emission spectroscopy. In principle, one could make an independent fit for each filter separately. For simplicity, we use WFC3 here in the following. To constrain the three parameters, we perform a global minimisation by using a three-dimensional Nelder-Mead Simplex Method \citep{LagariasReedsWrightEtAl98}. The quantity to be minimised is the mean error in the WFC3 filter, integrated over a range of particle sizes with radii between $a_l = 0.1 \micron$ and $a_u = 10 \micron$:
\begin{equation}
 \overline{E} = \frac{1}{\log a_u - \log a_l} \int_{\log a_l}^{\log a_u} E(a) \, \mathrm d \log a \ ,
\end{equation}
with the mean error $E$ in a wavelength range between $\lambda_l$ and $\lambda_u$ given by
\begin{equation}
  E(a) = \frac{1}{\lambda_u - \lambda_l} \int_{\lambda_l}^{\lambda_u} \left| Q_\text{ext,mie}(a, \lambda) - Q_\text{ext,fit}(a, \lambda) \right| \mathrm d \lambda \ .
  \label{eq:filter_error}
\end{equation}
The integration over the particle sizes in the calculation of $\overline{E}$ is performed over the logarithm of $a$ because the radii span two orders of magnitude. We limit the largest radii to $a_u = 10 \micron$ because larger particles would already be in the large particle limit at the small wavelengths under investigation here, i.e. their extinction efficiencies would yield $Q_{\text{ext}} = 2$, irrespective of the actual particle size. The efficiencies have been calculated with the Mie code described in Sect. \ref{sec:mie_theory} for a broad range different particle radii between $a_u = 0.1 \micron$ and $a_u = 10 \micron$. Examples of the resulting extinction efficiencies for three different sizes and three different condensates are shown in Fig. \ref{fig:fit_examples}.

\subsection{Fit results}
\label{sec:fit_results}

\begin{table}
	\caption{Coefficients of the analytic fits of the extinction efficiencies and their corresponding average errors. The errors shown in the table refer to the WFC3 filter The condensates are ordered with respect to their $Q_0$ value.}
	\label{tab:analytic_fits}
	\centering\begin{tabular}{lrrrr}
		\hline
		Condensate   & $Q_0$ & $Q_1$ & $a$ & $\overline{E}$ (\%)\\
		\hline
		
		\ce{C}                    & 0.07  & 3.62 & 6.58 & 8.95  \\ [2pt]   
		\ce{TiC}                  & 0.24  & 3.60 & 3.74 & 7.45  \\ [2pt]
		\ce{Cr}                   & 0.30  & 3.61 & 3.66 & 7.96  \\ [2pt]
		\ce{Fe}[s]                & 0.30  & 3.59 & 3.88 & 8.09  \\ [2pt] 		
		\ce{FeS}                  & 0.30  & 3.84 & 3.12 & 6.01  \\ [2pt]		
		\ce{FeO}                  & 0.49  & 3.78 & 4.31 & 7.78  \\ [2pt]
		\ce{Fe2O3}                & 0.99  & 3.81 & 5.05 & 12.38 \\ [2pt]
		\ce{MnS}                  & 1.11  & 3.84 & 5.44 & 15.35 \\ [2pt]
		\ce{SiC}                  & 1.47  & 3.86 & 5.23 & 13.19 \\ [2pt]
		\ce{TiO2}                 & 1.74  & 3.90 & 5.21 & 13.30 \\ [2pt]
		\ce{ZnS}                  & 2.16  & 3.88 & 4.96 & 12.80 \\ [2pt]	
		\ce{CaTiO3}               & 2.17  & 3.94 & 5.07 & 12.88 \\ [2pt]
		\ce{SiO}                  & 2.48  & 4.02 & 4.25 & 8.66  \\ [2pt]
		\ce{Mg_{0.8}Fe_{1.2}SiO4} & 3.56  & 4.00 & 3.55 & 10.05 \\ [2pt]
		\ce{MgFeSiO4}             & 4.68  & 3.99 & 3.53 & 10.45 \\ [2pt]
		\ce{Fe2SiO4}              & 5.52  & 4.09 & 4.38 & 11.28 \\ [2pt]
		\ce{Na2S}                 & 6.77  & 4.07 & 4.03 & 10.79 \\ [2pt]
		\ce{MgO}                  & 8.23  & 4.09 & 4.15 & 10.18 \\ [2pt]
		\ce{Mg_{0.4}Fe_{0.6}SiO3} & 8.51  & 4.09 & 3.92 & 9.93  \\ [2pt]
		\ce{MgAl2O4}              & 8.85  & 4.10 & 4.18 & 9.90  \\ [2pt]
		\ce{Mg_{0.5}Fe_{0.5}SiO3} & 9.54  & 4.10 & 3.96 & 9.57  \\ [2pt]
		\ce{Al2O3}                & 9.55  & 4.16 & 3.51 & 9.99  \\ [2pt] 
		Titan tholins             & 10.69 & 4.15 & 4.07 & 9.49  \\ [2pt]		
		\ce{Mg2SiO4}              & 11.95 & 4.16 & 4.05 & 9.29  \\ [2pt]
		\ce{Mg_{0.8}Fe_{0.2}SiO3} & 13.27 & 4.18 & 3.95 & 8.66  \\ [2pt]
		\ce{MgSiO3}$^{(a)}$       & 15.15 & 4.17 & 3.98 & 8.93  \\ [2pt]
		\ce{MgSiO3}$^{(b)}$       & 15.59 & 4.21 & 3.98 & 8.80  \\ [2pt]
		\ce{SiO2}$^{(c)}$         & 16.47 & 4.21 & 3.95 & 8.79  \\ [2pt]
		\ce{NaCl}                 & 16.59 & 4.22 & 3.95 & 8.74  \\ [2pt]	
		\ce{KCl}                  & 20.76 & 4.18 & 3.84 & 8.68  \\ [2pt]
		\ce{H2O}[s]               & 64.98 & 4.36 & 3.34 & 9.53  \\ [2pt]
		\hline
	\end{tabular}\\
	\vspace{1ex}
	\textbf{Notes.} $^{(a)}$ amorphous (glassy); $^{(b)}$ amorphous (sol-gel); \\ $^{(c)}$ $\alpha$-crystal \& amorphous 
\end{table}

\begin{figure}
	\includegraphics[width=\columnwidth]{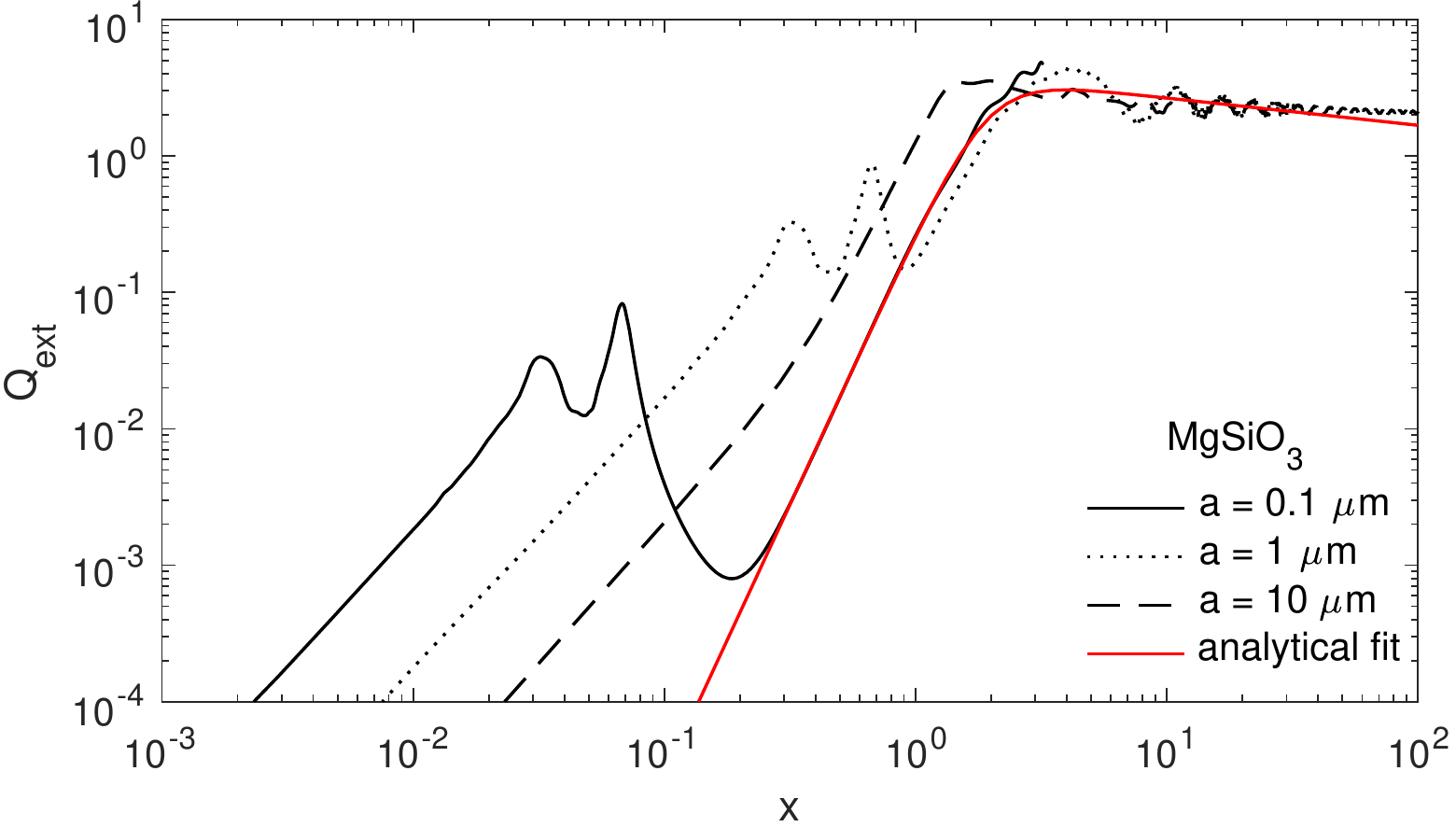}
	\includegraphics[width=\columnwidth]{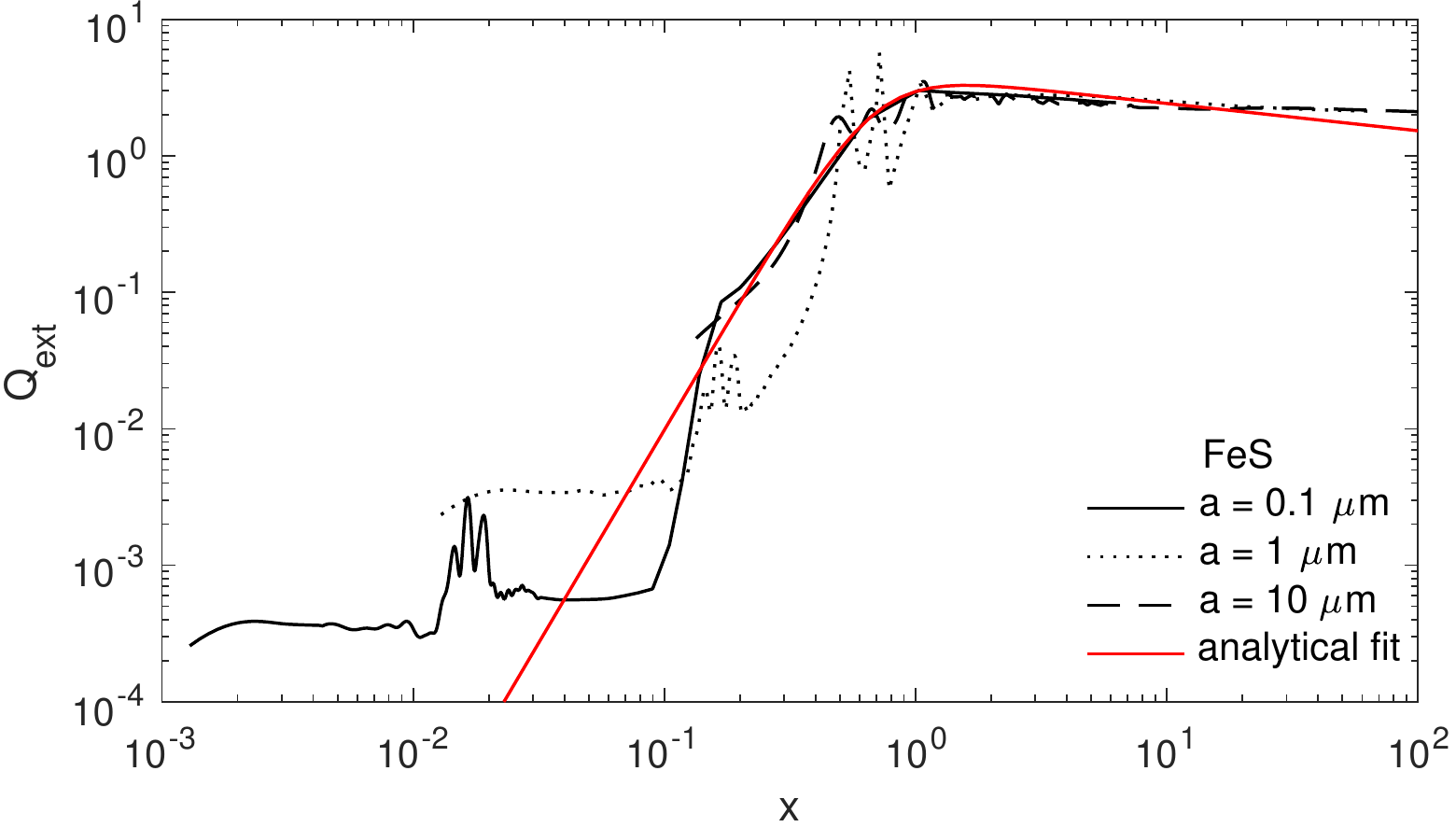}
	\includegraphics[width=\columnwidth]{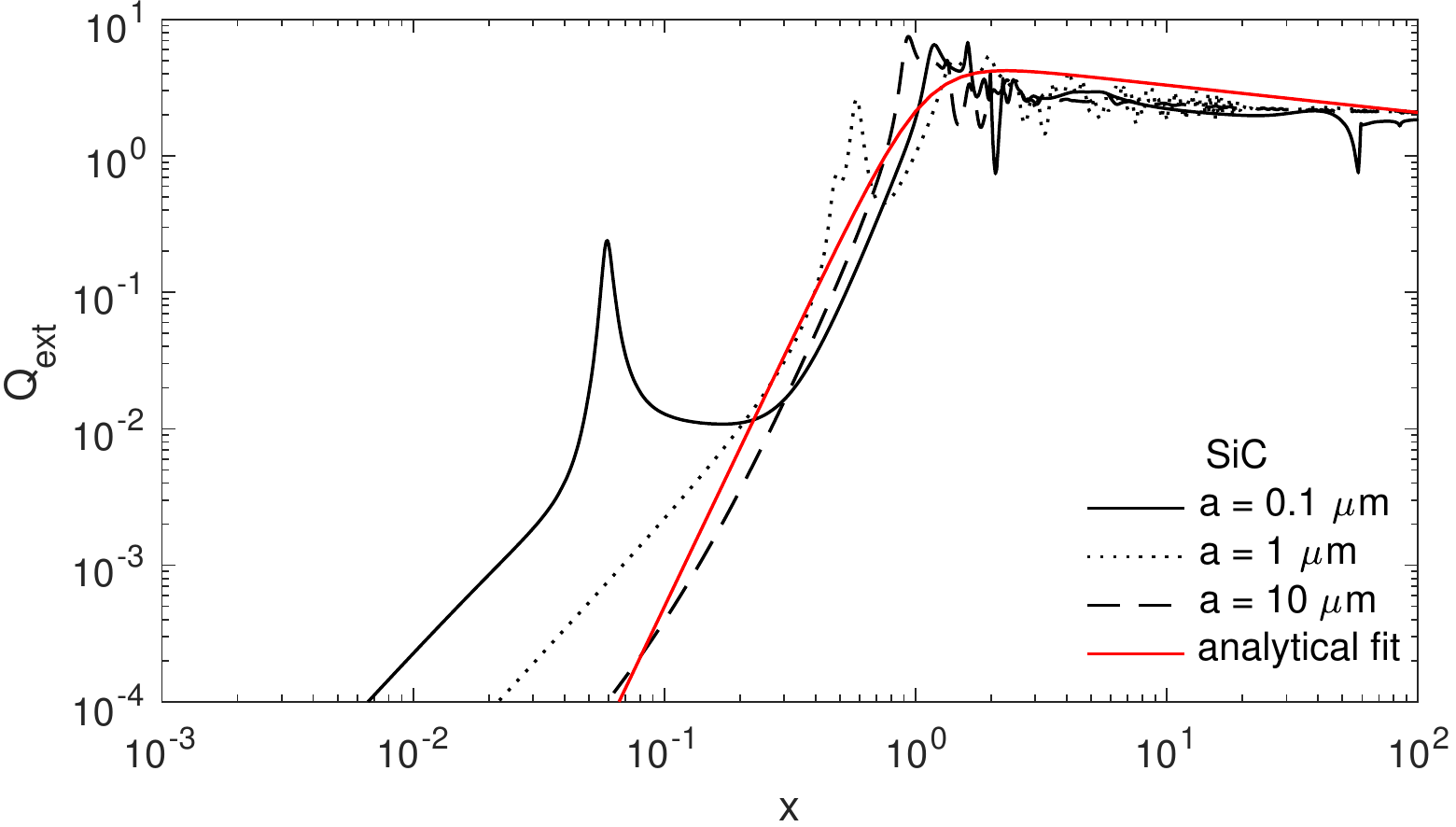}
	\caption{Examples of analytical fits for three different condensates. The analytical fits are depicted by the red, solid curves as a function of the size parameter $x$. For comparison, the $Q_\mathrm{ext}$ values are also shown in black for three different particles sizes:  0.1 $\micron$ (solid), 1 $\micron$ (dotted), and 10 $\micron$ (dashed).}
	\label{fig:fit_examples}
\end{figure}

The obtained fit coefficients and the mean error $\overline{E}$ are shown in Table \ref{tab:analytic_fits} in descending order of their $Q_0$ parameter. As mentioned in \citet{Lee2013ApJ...778...97L}, refractory condensates feature $Q_0$ values around 10-20, while more volatile species, such as water, have much higher ones. The $Q_0$ values of elements like iron, chromium, or graphite, on the other hand, drop below unity. Based on this range of values, it might be possible to identify condensates present in exoplanet atmosphere to a certain degree by employing retrieval techniques (see e.g. \citealt{Lavie2017AJ....154...91L}).

The normalisation constant $Q_1$ shows a much smaller spread, with values close to 4. The exponential parameter varies roughly between 3 and 7, depending on the absorptivity of the materials in the wavelength range under consideration. Thus, for actual retrieval methods, the normalisation constant $Q_1$ could be fixed to 4 in many cases since the accuracy of exoplanet data is currently not good enough to distinguish the very close parameter values for many of the condensates.

Examples of the analytical fits in comparison to the actual $Q_{\text{ext}}$ curves obtained from Mie theory are shown in Fig. \ref{fig:fit_examples} for three different condensates. The results suggest that the analytical function can represent the actual Mie efficiencies in certain wavelength ranges. Obviously, the simple fit cannot account for absorption features of the different condensates, as can be seen for the cases of \ce{FeS} or moissanite.

\begin{figure*}
	\includegraphics[width=155mm]{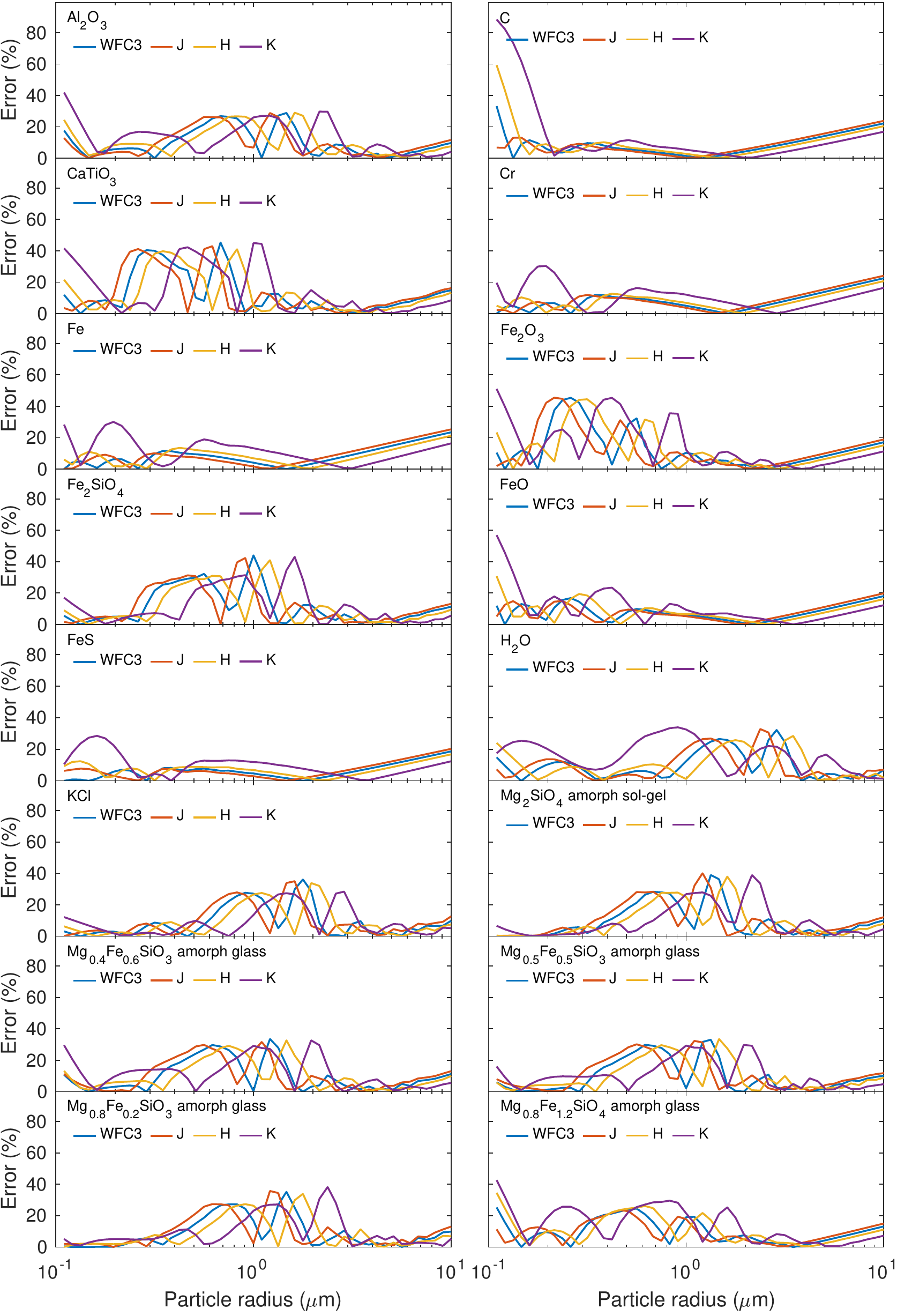}
\end{figure*}

\begin{figure*}
	\includegraphics[width=155mm]{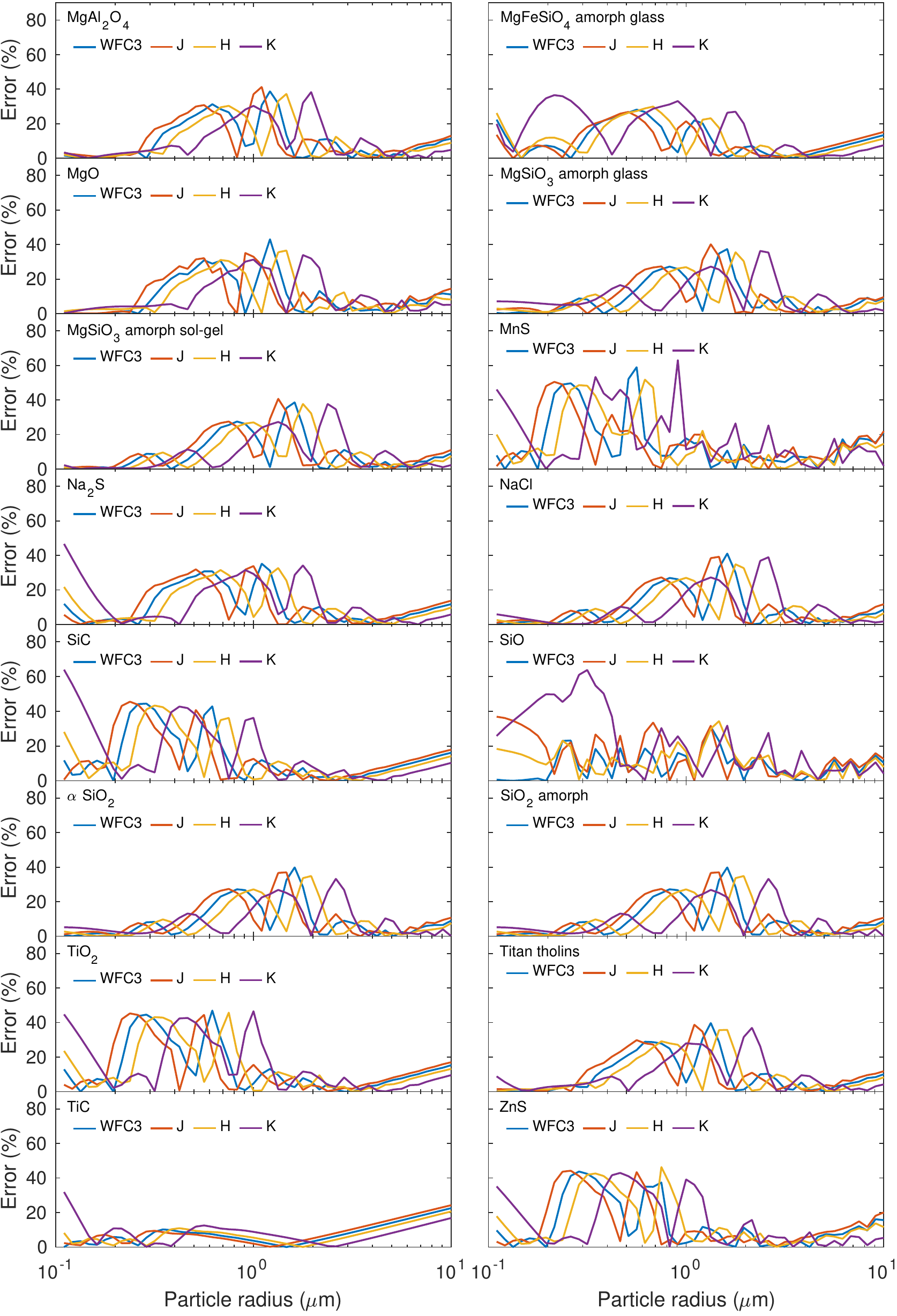}
	\caption{Errors of the analytical fits as a function of particle radius. The errors are shown for four different filters commonly used in exoplanet and brown dwarf spectroscopy: G141@WFC3 (blue), J (red), H (yellow), and K (purple).}
	\label{fig:fit_errors}
\end{figure*}

More detailed information on the errors of the fits as a function of particle sizes are shown in Fig. \ref{fig:fit_errors}. The figure depicts the mean errors, averaged over four different filters, as a function of the particle sizes. Equation \eqref{eq:filter_error} is used to calculate the mean error in each filter. The results shown in Fig. \ref{fig:fit_errors} suggest that depending on the type of condensate and the particle size, the errors can be as small as 1\%, but can also yield values around 80\%, especially for very small particle radii.

\section{Discussion and prospects for future work}
\label{sec:summary}

In this work, we compiled a new set of optical constants for potential condensates in atmospheres of giant exoplanets and brown dwarfs. While we tried to provide a consistent data set over a broad wavelength range for each species, the data needs to be treated with care as the lack of available laboratory data currently does not allow to provide more accurate refractive indices or for intercomparisons since in most cases only single measurements were available. Gaps in the experimental data needed to be closed by extrapolation of the $n$ or $k$ values and by employing the Kramers-Kronig relations. Other uncertainties for exoplanet atmospheres include also the effect of temperature or porosity on the optical constants.

Optical constants are clearly a function of temperature (see e.g. \citealt{Zeidler2013AA...553A..81Z}). Temperature-dependent effects include, for example, a shift and/or broadening of absorption features or changes in the general slopes of the $n$ and $k$ values. In most cases, however, the refractive index has been measured at room temperature or in environments resembling the cold interstellar medium. Data for temperatures more related to the atmospheres of brown dwarfs or extrasolar giant planets are rarely available and measurements for these cases are strongly required. Additionally, the optical constants of sulphide-bearing species need additional laboratory measurements to close the gaps in the present data, especially for \ce{MnS} and \ce{FeS}. Furthermore, most solid condensates forming in atmospheres are expected to be amorphous, which also directly impacts the refractive index. To date, unfortunately, this data is unavailable for many condensates, though the measurements by e.g. \citet{Dorschner1995AA...300..503D} or \citet{Jager2003AA...408..193J} are notable exceptions. The compilation of optical constants listed in Table \ref{tab:optical_constants} and shown in Fig. \ref{fig:optical_constants} is available as part of this publication's online material and on the Exoclime Simulation Platform (\url{http://www.exoclime.org}).

The optical properties of the condensates were calculated by using our newly developed, improved Mie-scattering code. It is able to calculate the Mie efficiencies, asymmetry parameters, and scattering phase functions for size parameters exceeding values of $10^7$. The code has been extensively tested against other available codes. It will be publicly released as part of the Exoclime Simulation Platform. With the calculated optical properties, we performed analytical fits of the obtained extinction efficiencies. The fits can be used in atmospheric retrieval of exoplanet atmospheres by adding two to three additional parameters to the set of retrievable quantities that describe the basic shapes of the extinction efficiencies. They, thus, allow for a basic characterisation of the clouds present in these atmospheres. We will employ these analytical fits for atmospheric retrieval in a future publication.

\section*{Acknowledgements}
The authors knowledge financial and administrative support by the Center for Space and Habitability and the PlanetS National Centre of Competence in Research (NCCR), as well as funding from the Swiss National Science Foundation (SNSF). D.K. would like to thank A.B.C. Patzer for helping to obtain the optical constants of \ce{TiC}.

\appendix
\section{Mie scattering code validation}
\label{sec:appendix_validation}

\begin{table*}
	\caption[]{Comparison of extinction and scattering efficiencies obtained from \citet{Du2004ApplOp}, \texttt{MIEV0} \citep{Wiscombe1979msca}, and \texttt{LX-MIE} (this work) for different refractive indices and size parameters.}
	\label{tab:validation}
	\centering
	\begin{tabular}{c c c c c c c c}
		\hline
		\noalign{\smallskip}
		& & \multicolumn{3}{c}{$Q_\mathrm{ext}$} & \multicolumn{3}{c}{$Q_\mathrm{sca}$}\\
		$m$ & $x$ & Du (2004) & \texttt{MIEV0} & \texttt{LX-MIE} & Du (2004) & \texttt{MIEV0} & \texttt{LX-MIE}\\
		\noalign{\smallskip}
		\hline
		\noalign{\smallskip}
		$10 - 10\mathrm{i}$ & $0.001$ & & $6.00207512 \times 10^{-5}$ & $6.00207581 \times 10^{-5}$ & & $2.66646989 \times 10^{-12}$ & $2.66646989 \times 10^{-12}$\\
		$10 - 10\mathrm{i}$ & $0.1$ & & $3.20390865 \times 10^{-2}$ & $3.20390907 \times 10^{-2}$ & & $2.70983004 \times 10^{-4}$ & $2.70983004 \times 10^{-4}$\\
		$10 - 10\mathrm{i}$ & $1$ & $2.53299$ & $2.53299308$ & $2.53299308$ & $2.04941$ & $2.04940501$ & $2.04940501$\\
		$10 - 10\mathrm{i}$ & $100$ & $2.07112$ & $2.07112433$ & $2.07112433$ & $1.83679$ & $1.83678540$ & $1.83678540$\\
		$10 - 10\mathrm{i}$ & $1000$ & & $2.02426046$ & $2.02426046$ & & $1.80546582$ & $1.80546582$\\
		$10 - 10\mathrm{i}$ & $20000$ & & $2.00361474$ & $2.00361474$ & & $1.79419080$ & $1.79419080$\\
		$10 - 10\mathrm{i}$ & $10^{6}$ & $2.00022$ & & $2.00021914$ & $1.79218$ & & $1.79218105$\\
		$1.5 - 1\mathrm{i}$ & $100$ & $2.09750$ & $2.09750176$ & $2.09750176$ & $1.28370$ & $1.28369705$ & $1.28369705$\\
		$1.5 - 1\mathrm{i}$ & $10000$ & $2.00437$ & $2.00436771$ & $2.00436771$ & $1.23657$ & $1.23657431$ & $1.23657431$\\
		$1.33 - 10^{-5} \mathrm{i}$ & $100$ & $2.10132$ & $2.10132071$ & $2.10132071$ & $2.09659$ & $2.09659351$ & $2.09659351$\\
		$1.33 - 10^{-5} \mathrm{i}$ & $10000$ & $2.00409$ & $2.00408893$ & $2.00408893$ & $1.72386$ & $1.72385722$ & $1.72385722$\\
		$0.75 - 0 \mathrm{i}$ & $10$ & $2.23226$ & $2.23226484$ & $2.23226484$ & $2.23226$ & $2.23226484$ & $2.23226484$\\
		$0.75 - 0 \mathrm{i}$ & $1000$ & $1.99791$ & $1.99790818$ & $1.99790818$ & $1.99791$ & $1.99790818$ & $1.99790818$\\ 	
		$0.75 - 0 \mathrm{i}$ & $10000$ & & $2.00125518$ & $2.00125518$ & & $2.00125518$ & $2.00125518$\\
		\noalign{\smallskip}
		\hline
	\end{tabular}
\end{table*}

The Mie code developed in this work has been extensively validated against the results of other standard Mie codes, such as the well-known code \texttt{MIEV0} \citep{Wiscombe1979msca}. In this appendix we present a sample of the validation results.

\begin{figure}
	\centering
	\resizebox{\hsize}{!}{\includegraphics{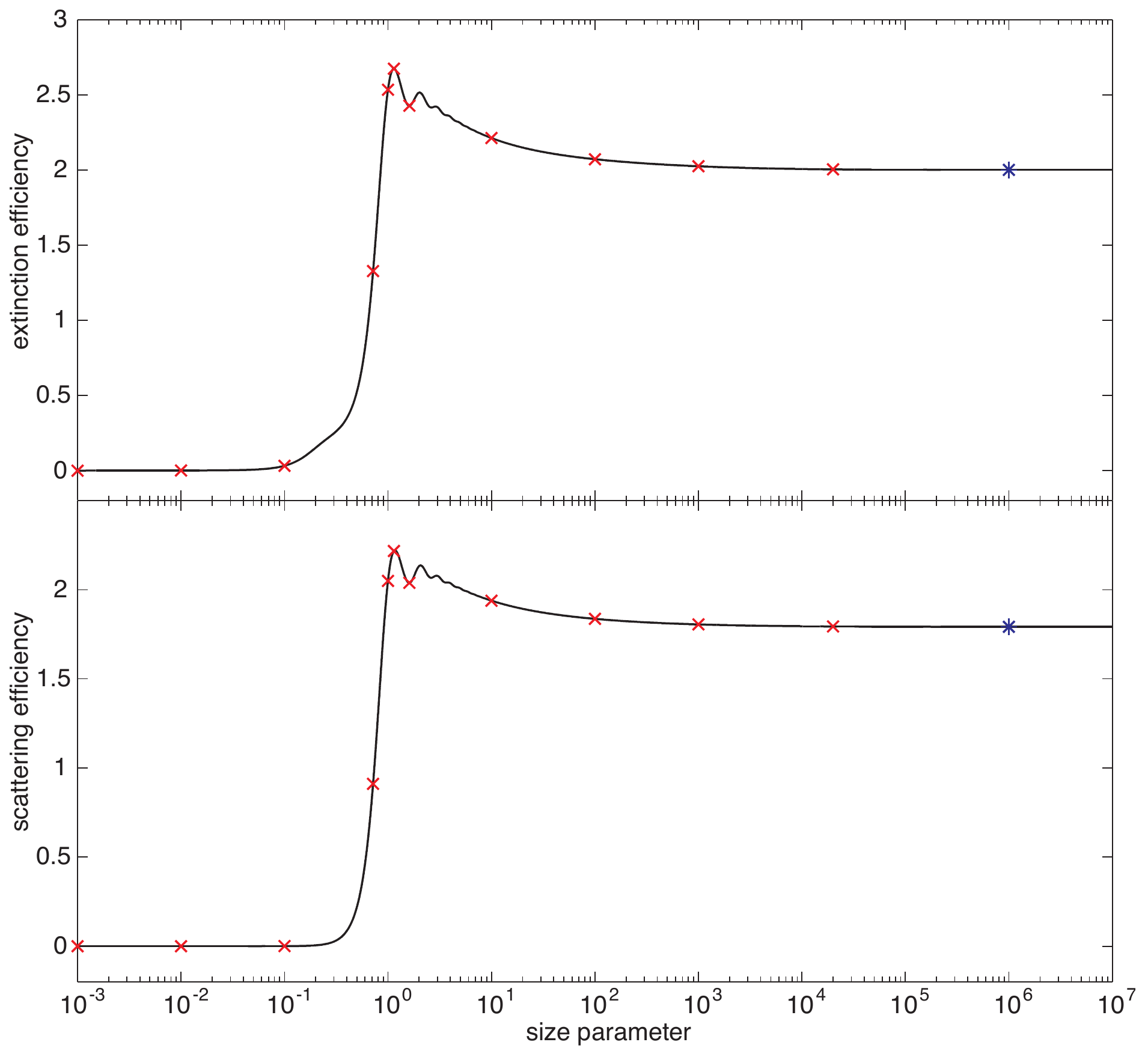}}
	\caption{Extinction efficiency (\textit{upper diagram}) and scattering efficiency (\textit{lower diagram}) obtained from Mie scattering calculations with \texttt{LX-MIE} for a refractive index of $m= 10 - 10\mathrm{i}$. Selected results of the \texttt{MIEV0} code from \citet{Wiscombe1979msca} are shown as red crosses for several selected representative size parameters. Additionally, the results for $x = 10^6$ published by \citet{Du2004ApplOp} are marked by blue stars.}
	\label{fig:validation}
\end{figure}

For the intercomparison, we use a set of refractive indices published by e.g. \citet{Du2004ApplOp} or \citet{Shen2005PIERS} which includes cases of strongly, weakly, and non-absorbing particles. The values for the extinction efficiency $Q_\mathrm{ext}$ and scattering efficiency $Q_\mathrm{sca}$ obtained from our code \texttt{LX-MIE}, \texttt{MIEV0} (in double-precision accuracy), and published results from \citet{Du2004ApplOp} are listed in Table \ref{tab:validation} for different refractive indices $m$ and several specific size parameters $x$. Additionally, the extinction and scattering efficiencies for a refractive index of $m= 10 - 10\mathrm{i}$ are shown in Fig. \ref{fig:validation} as a function of the size parameter $x$. Selected values for representative size parameters calculated with \texttt{MIEV0} are also shown in this figure within its validated range of $x < 20\,000$. For a quantitative comparison, some of these values are listed in Table \ref{tab:validation}.

The resulting efficiencies in Table \ref{tab:validation} and Fig. \ref{fig:validation} show that the results of our Mie code and \texttt{MIEV0} agree in at least $9$ significant digits in most cases. Some minor deviations are found for small size parameters, which are caused by the small-particle approximations used in \texttt{MIEV0} (see \citet{Wiscombe1979msca} for details on its treatment of Mie calculations for small size parameters). 
Note, that the results from \citet{Du2004ApplOp} listed in Table \ref{tab:validation} are identical to those published by \citet{Shen2005PIERS}, who introduced the improved treatment of the Riccati-Bessel functions which is now also used in \texttt{LX-MIE}.

For a large size parameter of $x = 10^6$, the values given in \citet{Du2004ApplOp} are used for comparison. His results of $Q_\mathrm{ext} = 2.00022$ and $Q_\mathrm{sca} = 1.79218$ agree with our calculated values in all digits. While $x = 10^6$ is the largest size parameter for which results were published by \citet{Du2004ApplOp}, Fig. \ref{fig:validation} suggests that \texttt{LX-MIE} yields numerically stable results for size parameters of at least $10^7$. Thus, \texttt{LX-MIE} provides accurate results over at least ten orders of magnitude in the size parameter range.

\bibliographystyle{mnras} 
\bibliography{references}

\begin{thebibliography}{}
\makeatletter
\relax
\def\mn@urlcharsother{\let\do\@makeother \do\$\do\&\do\#\do\^\do\_\do\%\do\~}
\def\mn@doi{\begingroup\mn@urlcharsother \@ifnextchar [ {\mn@doi@}
  {\mn@doi@[]}}
\def\mn@doi@[#1]#2{\def\@tempa{#1}\ifx\@tempa\@empty \href
  {http://dx.doi.org/#2} {doi:#2}\else \href {http://dx.doi.org/#2} {#1}\fi
  \endgroup}
\def\mn@eprint#1#2{\mn@eprint@#1:#2::\@nil}
\def\mn@eprint@arXiv#1{\href {http://arxiv.org/abs/#1} {{\tt arXiv:#1}}}
\def\mn@eprint@dblp#1{\href {http://dblp.uni-trier.de/rec/bibtex/#1.xml}
  {dblp:#1}}
\def\mn@eprint@#1:#2:#3:#4\@nil{\def\@tempa {#1}\def\@tempb {#2}\def\@tempc
  {#3}\ifx \@tempc \@empty \let \@tempc \@tempb \let \@tempb \@tempa \fi \ifx
  \@tempb \@empty \def\@tempb {arXiv}\fi \@ifundefined
  {mn@eprint@\@tempb}{\@tempb:\@tempc}{\expandafter \expandafter \csname
  mn@eprint@\@tempb\endcsname \expandafter{\@tempc}}}

\bibitem[\protect\citeauthoryear{{Abramowitz} \& {Stegun}}{{Abramowitz} \&
  {Stegun}}{1972}]{AbramowitzStegun1972}
{Abramowitz} M.,  {Stegun} I.~A.,  1972, {Handbook of Mathematical Functions}

\bibitem[\protect\citeauthoryear{{Arney} et~al.,}{{Arney}
  et~al.}{2016}]{Arney2016AsBio..16..873A}
{Arney} G.,  et~al., 2016, \mn@doi [Astrobiology] {10.1089/ast.2015.1422},
  \href {http://adsabs.harvard.edu/abs/2016AsBio..16..873A} {16, 873}

\bibitem[\protect\citeauthoryear{{Begemann}, {Dorschner}, {Henning},
  {Mutschke}, {G{\"u}rtler}, {K{\"o}mpe}  \& {Nass}}{{Begemann}
  et~al.}{1997}]{Begemann1997ApJ...476..199B}
{Begemann} B.,  {Dorschner} J.,  {Henning} T.,  {Mutschke} H.,  {G{\"u}rtler}
  J.,  {K{\"o}mpe} C.,   {Nass} R.,  1997, \mn@doi [\apj] {10.1086/303597},
  \href {http://adsabs.harvard.edu/abs/1997ApJ...476..199B} {476, 199}

\bibitem[\protect\citeauthoryear{{Birnstiel}, {Klahr}  \&
  {Ercolano}}{{Birnstiel} et~al.}{2012}]{Birnstiel2012A&A...539A.148B}
{Birnstiel} T.,  {Klahr} H.,   {Ercolano} B.,  2012, \mn@doi [\aap]
  {10.1051/0004-6361/201118136}, \href
  {http://adsabs.harvard.edu/abs/2012A%26A...539A.148B} {539, A148}

\bibitem[\protect\citeauthoryear{{Bohren} \& {Huffman}}{{Bohren} \&
  {Huffman}}{1998}]{BohrenHuffman1998asls}
{Bohren} C.~F.,  {Huffman} D.~R.,  1998, {Absorption and Scattering of Light by
  Small Particles}

\bibitem[\protect\citeauthoryear{Cachorro \& Salcedo}{Cachorro \&
  Salcedo}{1991}]{Cachorro1991}
Cachorro V.,  Salcedo L.,  1991, \mn@doi [Journal of Electromagnetic Waves and
  Applications] {10.1163/156939391X00950}, 5, 913

\bibitem[\protect\citeauthoryear{{Chandrasekhar}}{{Chandrasekhar}}{1960}]{Chandrasekhar1960ratr}
{Chandrasekhar} S.,  1960, {Radiative transfer}

\bibitem[\protect\citeauthoryear{{Colaprete} \& {Toon}}{{Colaprete} \&
  {Toon}}{2003}]{Colaprete2003}
{Colaprete} A.,  {Toon} O.~B.,  2003, \mn@doi [Journal of Geophysical Research
  (Planets)] {10.1029/2002JE001967}, \href
  {http://adsabs.harvard.edu/abs/2003JGRE..108.5025C} {108, 5025}

\bibitem[\protect\citeauthoryear{{Dave}}{{Dave}}{1970}]{Dave1970ApOpt}
{Dave} J.~V.,  1970, \ao, \href
  {http://adsabs.harvard.edu/abs/1970ApOpt...9.1888D} {9, 1888}

\bibitem[\protect\citeauthoryear{{Dorschner}, {Begemann}, {Henning}, {Jaeger}
  \& {Mutschke}}{{Dorschner} et~al.}{1995}]{Dorschner1995AA...300..503D}
{Dorschner} J.,  {Begemann} B.,  {Henning} T.,  {Jaeger} C.,   {Mutschke} H.,
  1995, \aap, \href {http://adsabs.harvard.edu/abs/1995A%26A...300..503D} {300,
  503}

\bibitem[\protect\citeauthoryear{{Draine}}{{Draine}}{2003}]{Draine2003ApJ...598.1017D}
{Draine} B.~T.,  2003, \mn@doi [\apj] {10.1086/379118}, \href
  {http://adsabs.harvard.edu/abs/2003ApJ...598.1017D} {598, 1017}

\bibitem[\protect\citeauthoryear{{Du}}{{Du}}{2004}]{Du2004ApplOp}
{Du} H.,  2004, \mn@doi [Applied Optics] {10.1364/AO.43.001951}, 43, 1951

\bibitem[\protect\citeauthoryear{{Fabian}, {Henning}, {J{\"a}ger}, {Mutschke},
  {Dorschner}  \& {Wehrhan}}{{Fabian} et~al.}{2001}]{Fabian2001AA...378..228F}
{Fabian} D.,  {Henning} T.,  {J{\"a}ger} C.,  {Mutschke} H.,  {Dorschner} J.,
  {Wehrhan} O.,  2001, \mn@doi [\aap] {10.1051/0004-6361:20011196}, \href
  {http://adsabs.harvard.edu/abs/2001A%26A...378..228F} {378, 228}

\bibitem[\protect\citeauthoryear{{Helling}, {Woitke}  \& {Thi}}{{Helling}
  et~al.}{2008}]{Helling2008A&A...485..547H}
{Helling} C.,  {Woitke} P.,   {Thi} W.-F.,  2008, \mn@doi [\aap]
  {10.1051/0004-6361:20078220}, \href
  {http://adsabs.harvard.edu/abs/2008A\%26A...485..547H} {485, 547}

\bibitem[\protect\citeauthoryear{{Heng}}{{Heng}}{2016}]{Heng2016ApJ...826L..16H}
{Heng} K.,  2016, \mn@doi [\apjl] {10.3847/2041-8205/826/1/L16}, \href
  {http://adsabs.harvard.edu/abs/2016ApJ...826L..16H} {826, L16}

\bibitem[\protect\citeauthoryear{{Henning} \& {Mutschke}}{{Henning} \&
  {Mutschke}}{1997}]{Henning1997AA...327..743H}
{Henning} T.,  {Mutschke} H.,  1997, \aap, \href
  {http://adsabs.harvard.edu/abs/1997A%26A...327..743H} {327, 743}

\bibitem[\protect\citeauthoryear{{Henning} \& {Mutschke}}{{Henning} \&
  {Mutschke}}{2001}]{Henning2001AcSpe..57..815H}
{Henning} T.,  {Mutschke} H.,  2001, Spectrochimica Acta, \href
  {http://adsabs.harvard.edu/abs/2001AcSpe..57..815H} {57, 815}

\bibitem[\protect\citeauthoryear{{Henning}, {Begemann}, {Mutschke}  \&
  {Dorschner}}{{Henning} et~al.}{1995}]{Henning1995AAS..112..143H}
{Henning} T.,  {Begemann} B.,  {Mutschke} H.,   {Dorschner} J.,  1995, \aaps,
  \href {http://adsabs.harvard.edu/abs/1995A%26AS..112..143H} {112, 143}

\bibitem[\protect\citeauthoryear{{Huffman} \& {Wild}}{{Huffman} \&
  {Wild}}{1967}]{Huffman1967PhRv..156..989H}
{Huffman} D.~R.,  {Wild} R.~L.,  1967, \mn@doi [Physical Review]
  {10.1103/PhysRev.156.989}, \href
  {http://adsabs.harvard.edu/abs/1967PhRv..156..989H} {156, 989}

\bibitem[\protect\citeauthoryear{{Infeld}}{{Infeld}}{1947}]{Infeld1947QAM}
{Infeld} L.,  1947, Quart. Appl. Math., 5, 113

\bibitem[\protect\citeauthoryear{{J{\"a}ger}, {Dorschner}, {Mutschke}, {Posch}
  \& {Henning}}{{J{\"a}ger} et~al.}{2003}]{Jager2003AA...408..193J}
{J{\"a}ger} C.,  {Dorschner} J.,  {Mutschke} H.,  {Posch} T.,   {Henning} T.,
  2003, \mn@doi [\aap] {10.1051/0004-6361:20030916}, \href
  {http://adsabs.harvard.edu/abs/2003A%26A...408..193J} {408, 193}

\bibitem[\protect\citeauthoryear{Kangarloo \& Rafizadeh}{Kangarloo \&
  Rafizadeh}{2010a}]{kangarloo2010a}
Kangarloo H.,  Rafizadeh S.,  2010a, in Mechanical and Electronics Engineering
  (ICMEE), 2010 2nd International Conference on Mechanical and Electronics
  Engineering. pp V2--136

\bibitem[\protect\citeauthoryear{Kangarloo \& Rafizadeh}{Kangarloo \&
  Rafizadeh}{2010b}]{kangarloo2010b}
Kangarloo H.,  Rafizadeh S.,  2010b, in Mechanical and Electronics Engineering
  (ICMEE), 2010 2nd International Conference on Mechanical and Electronics
  Engineering. pp V2--237

\bibitem[\protect\citeauthoryear{{Khachai}, {Khenata}, {Bouhemadou}, {Haddou},
  {Reshak}, {Amrani}, {Rached}  \& {Soudini}}{{Khachai}
  et~al.}{2009}]{Khachai2009JPCM...21i5404K}
{Khachai} H.,  {Khenata} R.,  {Bouhemadou} A.,  {Haddou} A.,  {Reshak} A.~H.,
  {Amrani} B.,  {Rached} D.,   {Soudini} B.,  2009, \mn@doi [Journal of Physics
  Condensed Matter] {10.1088/0953-8984/21/9/095404}, \href
  {http://adsabs.harvard.edu/abs/2009JPCM...21i5404K} {21, 095404}

\bibitem[\protect\citeauthoryear{{Khare}, {Sagan}, {Arakawa}, {Suits},
  {Callcott}  \& {Williams}}{{Khare} et~al.}{1984}]{Khare1984Icar...60..127K}
{Khare} B.~N.,  {Sagan} C.,  {Arakawa} E.~T.,  {Suits} F.,  {Callcott} T.~A.,
  {Williams} M.~W.,  1984, \mn@doi [\icarus] {10.1016/0019-1035(84)90142-8},
  \href {http://adsabs.harvard.edu/abs/1984Icar...60..127K} {60, 127}

\bibitem[\protect\citeauthoryear{{Kitzmann}, {Patzer}, {von Paris}, {Godolt},
  {Stracke}, {Gebauer}, {Grenfell}  \& {Rauer}}{{Kitzmann}
  et~al.}{2010}]{Kitzmann2010A&A...511A..66K}
{Kitzmann} D.,  {Patzer} A.~B.~C.,  {von Paris} P.,  {Godolt} M.,  {Stracke}
  B.,  {Gebauer} S.,  {Grenfell} J.~L.,   {Rauer} H.,  2010, \mn@doi [\aap]
  {10.1051/0004-6361/200913491}, \href
  {http://adsabs.harvard.edu/abs/2010A\%26A...511A..66K} {511, A66}

\bibitem[\protect\citeauthoryear{{Koide}, {Shidara}, {Fukutani}, {Fujimori},
  {Miyahara}, {Kato}, {Otani}  \& {Ishizawa}}{{Koide}
  et~al.}{1990}]{Koide1990PhRvB..42.4979K}
{Koide} T.,  {Shidara} T.,  {Fukutani} H.,  {Fujimori} A.,  {Miyahara} T.,
  {Kato} H.,  {Otani} S.,   {Ishizawa} Y.,  1990, \mn@doi [\prb]
  {10.1103/PhysRevB.42.4979}, \href
  {http://adsabs.harvard.edu/abs/1990PhRvB..42.4979K} {42, 4979}

\bibitem[\protect\citeauthoryear{{Koike}, {Kaito}, {Yamamoto}, {Shibai},
  {Kimura}  \& {Suto}}{{Koike} et~al.}{1995}]{Koike1995Icar..114..203K}
{Koike} C.,  {Kaito} C.,  {Yamamoto} T.,  {Shibai} H.,  {Kimura} S.,   {Suto}
  H.,  1995, \mn@doi [\icarus] {10.1006/icar.1995.1055}, \href
  {http://adsabs.harvard.edu/abs/1995Icar..114..203K} {114, 203}

\bibitem[\protect\citeauthoryear{{Kramers}}{{Kramers}}{1927}]{Kramers1927}
{Kramers} H.~A.,  1927, Atti Cong. Intern. Fisica, 2, 545

\bibitem[\protect\citeauthoryear{{Kronig}}{{Kronig}}{1926}]{Kronig1926JOSA}
{Kronig} R.~D.~L.,  1926, Journal of the Optical Society of America
  (1917-1983), \href {http://adsabs.harvard.edu/abs/1926JOSA...12..547K} {12,
  547}

\bibitem[\protect\citeauthoryear{Lagarias, Reeds, Wright  \& Wright}{Lagarias
  et~al.}{1998}]{LagariasReedsWrightEtAl98}
Lagarias J.~C.,  Reeds J.~A.,  Wright M.~H.,   Wright P.~E.,  1998, SIAM
  Journal of Optimization, 9, 112

\bibitem[\protect\citeauthoryear{{Laor} \& {Draine}}{{Laor} \&
  {Draine}}{1993}]{Laor1993ApJ...402..441L}
{Laor} A.,  {Draine} B.~T.,  1993, \mn@doi [\apj] {10.1086/172149}, \href
  {http://adsabs.harvard.edu/abs/1993ApJ...402..441L} {402, 441}

\bibitem[\protect\citeauthoryear{{Lavie} et~al.,}{{Lavie}
  et~al.}{2017}]{Lavie2017AJ....154...91L}
{Lavie} B.,  et~al., 2017, \mn@doi [\aj] {10.3847/1538-3881/aa7ed8}, \href
  {http://adsabs.harvard.edu/abs/2017AJ....154...91L} {154, 91}

\bibitem[\protect\citeauthoryear{{Lee}, {Heng}  \& {Irwin}}{{Lee}
  et~al.}{2013}]{Lee2013ApJ...778...97L}
{Lee} J.-M.,  {Heng} K.,   {Irwin} P.~G.~J.,  2013, \mn@doi [\apj]
  {10.1088/0004-637X/778/2/97}, \href
  {http://adsabs.harvard.edu/abs/2013ApJ...778...97L} {778, 97}

\bibitem[\protect\citeauthoryear{{Lentz}}{{Lentz}}{1976}]{Lentz1976ApOpt}
{Lentz} W.~J.,  1976, \mn@doi [\ao] {10.1364/AO.15.000668}, \href
  {http://adsabs.harvard.edu/abs/1976ApOpt..15..668L} {15, 668}

\bibitem[\protect\citeauthoryear{{Lucarini}, {Peiponen}, {Saarinen}  \&
  {Vartiainen}}{{Lucarini} et~al.}{2005}]{Lucarini2005kkro.book.....L}
{Lucarini} V.,  {Peiponen} K.-E.,  {Saarinen} J.~J.,   {Vartiainen} E.~M.,
  2005, {Kramers-Kronig Relations in Optical Materials Research},
  \mn@doi{10.1007/b138913.
}

\bibitem[\protect\citeauthoryear{{Marley}, {Ackerman}, {Cuzzi}  \&
  {Kitzmann}}{{Marley} et~al.}{2013}]{Marley2013cctp.book..367M}
{Marley} M.~S.,  {Ackerman} A.~S.,  {Cuzzi} J.~N.,   {Kitzmann} D.,  2013,
  {Clouds and Hazes in Exoplanet Atmospheres}.
pp 367--391, \mn@doi{10.2458/azu_uapress_9780816530595-ch15}

\bibitem[\protect\citeauthoryear{{Mie}}{{Mie}}{1908}]{Mie1908AnP}
{Mie} G.,  1908, \mn@doi [Annalen der Physik] {10.1002/andp.19083300302}, \href
  {http://adsabs.harvard.edu/abs/1908AnP...330..377M} {330, 377}

\bibitem[\protect\citeauthoryear{{Montaner}, {Galtier}, {Benoit}  \&
  {Bill}}{{Montaner} et~al.}{1979}]{Montaner1979PSSAR..52..597M}
{Montaner} A.,  {Galtier} M.,  {Benoit} C.,   {Bill} H.,  1979, \mn@doi
  [Physica Status Solidi Applied Research] {10.1002/pssa.2210520228}, \href
  {http://adsabs.harvard.edu/abs/1979PSSAR..52..597M} {52, 597}

\bibitem[\protect\citeauthoryear{{Mordasini}}{{Mordasini}}{2014}]{Mordasini2014A&A...572A.118M}
{Mordasini} C.,  2014, \mn@doi [\aap] {10.1051/0004-6361/201423702}, \href
  {http://adsabs.harvard.edu/abs/2014A%26A...572A.118M} {572, A118}

\bibitem[\protect\citeauthoryear{{Morley}, {Fortney}, {Marley}, {Visscher},
  {Saumon}  \& {Leggett}}{{Morley} et~al.}{2012}]{Morley2012ApJ...756..172M}
{Morley} C.~V.,  {Fortney} J.~J.,  {Marley} M.~S.,  {Visscher} C.,  {Saumon}
  D.,   {Leggett} S.~K.,  2012, \mn@doi [\apj] {10.1088/0004-637X/756/2/172},
  \href {http://adsabs.harvard.edu/abs/2012ApJ...756..172M} {756, 172}

\bibitem[\protect\citeauthoryear{{Oreshenko} et~al.,}{{Oreshenko}
  et~al.}{2017}]{Oreshenko2017ApJ...847L...3O}
{Oreshenko} M.,  et~al., 2017, \mn@doi [\apjl] {10.3847/2041-8213/aa8acf},
  \href {http://adsabs.harvard.edu/abs/2017ApJ...847L...3O} {847, L3}

\bibitem[\protect\citeauthoryear{{Palik}}{{Palik}}{1985}]{Palik1985hocs.book}
{Palik} E.~D.,  1985, {Handbook of optical constants of solids}

\bibitem[\protect\citeauthoryear{{Palik}}{{Palik}}{1991}]{Palik1991hocs.book}
{Palik} E.~D.,  1991, {Handbook of optical constants of solids II}

\bibitem[\protect\citeauthoryear{{Pierrehumbert}}{{Pierrehumbert}}{2010}]{Pierrehumbert2010ppc..book.....P}
{Pierrehumbert} R.~T.,  2010, {Principles of Planetary Climate}

\bibitem[\protect\citeauthoryear{{Pinhas} \& {Madhusudhan}}{{Pinhas} \&
  {Madhusudhan}}{2017}]{Pinhas2017MNRAS.471.4355P}
{Pinhas} A.,  {Madhusudhan} N.,  2017, \mn@doi [\mnras]
  {10.1093/mnras/stx1849}, \href
  {http://adsabs.harvard.edu/abs/2017MNRAS.471.4355P} {471, 4355}

\bibitem[\protect\citeauthoryear{{Pollack}, {Hollenbach}, {Beckwith},
  {Simonelli}, {Roush}  \& {Fong}}{{Pollack}
  et~al.}{1994}]{Pollack1994ApJ...421..615P}
{Pollack} J.~B.,  {Hollenbach} D.,  {Beckwith} S.,  {Simonelli} D.~P.,  {Roush}
  T.,   {Fong} W.,  1994, \mn@doi [\apj] {10.1086/173677}, \href
  {http://adsabs.harvard.edu/abs/1994ApJ...421..615P} {421, 615}

\bibitem[\protect\citeauthoryear{{Pont}, {Knutson}, {Gilliland}, {Moutou}  \&
  {Charbonneau}}{{Pont} et~al.}{2008}]{Pont2008MNRAS.385..109P}
{Pont} F.,  {Knutson} H.,  {Gilliland} R.~L.,  {Moutou} C.,   {Charbonneau} D.,
   2008, \mn@doi [\mnras] {10.1111/j.1365-2966.2008.12852.x}, \href
  {http://adsabs.harvard.edu/abs/2008MNRAS.385..109P} {385, 109}

\bibitem[\protect\citeauthoryear{{Posch}, {Kerschbaum}, {Fabian}, {Mutschke},
  {Dorschner}, {Tamanai}  \& {Henning}}{{Posch}
  et~al.}{2003}]{Posch2003ApJS..149..437P}
{Posch} T.,  {Kerschbaum} F.,  {Fabian} D.,  {Mutschke} H.,  {Dorschner} J.,
  {Tamanai} A.,   {Henning} T.,  2003, \mn@doi [\apjs] {10.1086/379167}, \href
  {http://adsabs.harvard.edu/abs/2003ApJS..149..437P} {149, 437}

\bibitem[\protect\citeauthoryear{Raki\'{c}, Djuri\v{s}i\'{c}, Elazar  \&
  Majewski}{Raki\'{c} et~al.}{1998}]{Rakic:98}
Raki\'{c} A.~D.,  Djuri\v{s}i\'{c} A.~B.,  Elazar J.~M.,   Majewski M.~L.,
  1998, \mn@doi [Appl. Opt.] {10.1364/AO.37.005271}, 37, 5271

\bibitem[\protect\citeauthoryear{{Rother}}{{Rother}}{2009}]{Rother2009ewsn.book.....R}
{Rother} T.,  2009, {Electromagnetic Wave Scattering on Nonspherical
  Particles}, \mn@doi{10.1007/978-3-642-00704-0.
}

\bibitem[\protect\citeauthoryear{Schmitt, Heymsfield, Connolly, Järvinen  \&
  Schnaiter}{Schmitt et~al.}{2016}]{SchmittGRL:GRL55211}
Schmitt C.~G.,  Heymsfield A.~J.,  Connolly P.,  Järvinen E.,   Schnaiter M.,
  2016, \mn@doi [Geophysical Research Letters] {10.1002/2016GL071267}, 43,
  11,913

\bibitem[\protect\citeauthoryear{{Shen} \& {Cai}}{{Shen} \&
  {Cai}}{2005}]{Shen2005PIERS}
{Shen} J.,  {Cai} X.,  2005, in {Progress in Electromagnetics Research
  Symposium}. pp 691--694

\bibitem[\protect\citeauthoryear{Siefke et~al.,}{Siefke
  et~al.}{2016}]{SiefkeADOM201600250}
Siefke T.,  et~al., 2016, \mn@doi [Advanced Optical Materials]
  {10.1002/adom.201600250}, 4, 1780

\bibitem[\protect\citeauthoryear{{Sing} et~al.,}{{Sing}
  et~al.}{2016}]{Sing2016Natur.529...59S}
{Sing} D.~K.,  et~al., 2016, \mn@doi [\nat] {10.1038/nature16068}, \href
  {http://adsabs.harvard.edu/abs/2016Natur.529...59S} {529, 59}

\bibitem[\protect\citeauthoryear{{Stevenson}, {Bean}, {Seifahrt}, {Gilbert},
  {Line}, {D{\'e}sert}  \& {Fortney}}{{Stevenson}
  et~al.}{2016}]{Stevenson2016ApJ...817..141S}
{Stevenson} K.~B.,  {Bean} J.~L.,  {Seifahrt} A.,  {Gilbert} G.~J.,  {Line}
  M.~R.,  {D{\'e}sert} J.-M.,   {Fortney} J.~J.,  2016, \mn@doi [\apj]
  {10.3847/0004-637X/817/2/141}, \href
  {http://adsabs.harvard.edu/abs/2016ApJ...817..141S} {817, 141}

\bibitem[\protect\citeauthoryear{{Tsiaras} et~al.,}{{Tsiaras}
  et~al.}{2017}]{Tsiaras2017arXiv170405413T}
{Tsiaras} A.,  et~al., 2017, preprint, \href
  {http://adsabs.harvard.edu/abs/2017arXiv170405413T} {} (\mn@eprint {arXiv}
  {1704.05413v1})

\bibitem[\protect\citeauthoryear{{Ueda}, {Yanagi}, {Noshiro}, {Hosono}  \&
  {Kawazoe}}{{Ueda} et~al.}{1998}]{Ueda1998JPCM...10.3669U}
{Ueda} K.,  {Yanagi} H.,  {Noshiro} R.,  {Hosono} H.,   {Kawazoe} H.,  1998,
  \mn@doi [Journal of Physics Condensed Matter] {10.1088/0953-8984/10/16/018},
  \href {http://adsabs.harvard.edu/abs/1998JPCM...10.3669U} {10, 3669}

\bibitem[\protect\citeauthoryear{{Wakeford} \& {Sing}}{{Wakeford} \&
  {Sing}}{2015}]{Wakeford2015A&A...573A.122W}
{Wakeford} H.~R.,  {Sing} D.~K.,  2015, \mn@doi [\aap]
  {10.1051/0004-6361/201424207}, \href
  {http://adsabs.harvard.edu/abs/2015A%26A...573A.122W} {573, A122}

\bibitem[\protect\citeauthoryear{{Warren}}{{Warren}}{1984}]{Warren1984ApOpt..23.1206W}
{Warren} S.~G.,  1984, \mn@doi [\ao] {10.1364/AO.23.001206}, \href
  {http://adsabs.harvard.edu/abs/1984ApOpt..23.1206W} {23, 1206}

\bibitem[\protect\citeauthoryear{{Wiscombe}}{{Wiscombe}}{1979}]{Wiscombe1979msca}
{Wiscombe} W.~J.,  1979, Technical report, {MIE scattering calculations,
  advances in technique and fast, vector-shaped computer codes}

\bibitem[\protect\citeauthoryear{{Wiscombe}}{{Wiscombe}}{1980}]{Wiscombe1980ApOpt}
{Wiscombe} W.~J.,  1980, \mn@doi [\ao] {10.1364/AO.19.001505}, \href
  {http://adsabs.harvard.edu/abs/1980ApOpt..19.1505W} {19, 1505}

\bibitem[\protect\citeauthoryear{{Zeidler}, {Posch}, {Mutschke}, {Richter}  \&
  {Wehrhan}}{{Zeidler} et~al.}{2011}]{Zeidler2011AA...526A..68Z}
{Zeidler} S.,  {Posch} T.,  {Mutschke} H.,  {Richter} H.,   {Wehrhan} O.,
  2011, \mn@doi [\aap] {10.1051/0004-6361/201015219}, \href
  {http://adsabs.harvard.edu/abs/2011A%26A...526A..68Z} {526, A68}

\bibitem[\protect\citeauthoryear{{Zeidler}, {Posch}  \& {Mutschke}}{{Zeidler}
  et~al.}{2013}]{Zeidler2013AA...553A..81Z}
{Zeidler} S.,  {Posch} T.,   {Mutschke} H.,  2013, \mn@doi [\aap]
  {10.1051/0004-6361/201220459}, \href
  {http://adsabs.harvard.edu/abs/2013A%26A...553A..81Z} {553, A81}

\bibitem[\protect\citeauthoryear{{van de Hulst}}{{van de
  Hulst}}{1957}]{Hulst1957lssp}
{van de Hulst} H.~C.,  1957, {Light Scattering by Small Particles}

\makeatother
\end{thebibliography}

\bsp	
\label{lastpage}
\end{document}